\documentclass[aps,showpacs,twocolumn,longbibliography,nofootinbib,secnumarabic,eqsecnum,superscriptaddress]{revtex4-1}

\usepackage{graphicx}
\usepackage{dcolumn}
\usepackage{bm}
\usepackage{braket}
\usepackage{leftidx}
\usepackage{cancel}
\usepackage{amsmath}
\usepackage{epstopdf}
\usepackage{color}
\usepackage[mathcal]{eucal}
\usepackage{float}

\usepackage{yfonts}
\usepackage{color}

\usepackage{url}

\usepackage{scalerel}

\usepackage{calligra}

\usepackage{pbsi}
\usepackage[T1]{fontenc}

\begin{document}
\date{\today}
\title{Nonclassicality and entanglement for  wavepackets}

\author{Mehmet Emre Tasgin}
\email{metasgin@hacettepe.edu.tr}
\affiliation{Institute of Nuclear Sciences, Hacettepe University, 06800, Ankara, Turkey}

\author{Mehmet Gunay}
\affiliation{Institute of Nuclear Sciences, Hacettepe University, 06800, Ankara, Turkey}

\author{M. Suhail Zubairy}
\affiliation{Institute of Quantum Studies and Department of Physics, Texas A {\&} M University, College Station, TX
77 843-4242, USA}

\begin{abstract}
Mode-entanglement based criteria and measures become insufficient for broadband emission, e.g. from spasers (plasmonic nano-lasers). We introduce criteria and measures for the (i) total entanglement of two wavepackets, (ii) entanglement of a wavepacket with an ensemble and (iii) total nonclassicality of a wavepacket~(WP). We discuss these criteria in the context of (i) entanglement of two WPs emitted from two initially entangled cavities (or two initially entangled atoms) and (ii) entanglement of an emitted WP with the ensemble/atom for the spontaneous emission and the single-photon superradiance. We also show that, (iii) when the two constituent modes of a WP are entangled, this creates nonclassicality in the WP as a noise reduction below the standard quantum limit. The criteria we introduce are, all, compatible with near-field detectors.
\end{abstract}

\maketitle

\section{Introduction}

Quantum entanglement, once appeared as a science-fiction phenomenon, became easily observable both in the macroscopic~\cite{QuantumTeleportation143km} and microscopic scales~\cite{20qubitsPRX2018}. Achievements like quantum teleportation~\cite{BraunsteinNatureTeleportation2015} with satellites~\cite{SatelliteTeleportation2017} or detection of stealth jets~\cite{QuantumRdarNews} with entangled microwave photons~\cite{QuantumRadarPRL2015} (quantum radars) made also the non-scientific community become aware of the importance of nonclassical phenomena which certainly will revolutionize the current technology. This makes the generation, detection and quantification of nonclassical states ---such as quadrature/number-squeezed\cite{PeggJModOpt1990Intelligent}, two-mode entangled and many-particle entangled states~\cite{duan2011many_particle_entanglement,tasgin2017many}--- much more important than the past century.

The last two decades witnessed stunning progress also in another research field: plasmonics and quantum plasmonics. Plasmonics affected all the fields of  science from sub-wavelength imaging of surfaces~(SNOM)~\cite{SNOMReview2006,SNOMNature2014} to sub-nm imaging of a molecule~\cite{SERS_subnmNature2013} and Raman-selective detection of ingredients via surface enhanced Raman scattering~(SERS)~\cite{SERS_bioanalysis_Review_NanoLett_2018}. Observation of phenomena analogous to electromagnetically induced transparency~(EIT)~\cite{ScullyZubairyBook} via path interference effects, e.g. Fano resonances~\cite{Fano_resonances_in_photonics_NaturePhot_2017,stockman_Nature_2010_Dark_hot_resonances,TasginFanoBook2018} and nonlinearity enhancement~\cite{PNAS_2013_Fano_FWM,SciRep_2016_Fano_CARS,tasgin_Nanophotonics_2018_SilentSERS}, made plasmonic systems more attractive. While plasmons decay much faster ($\tau_p\sim 10^{-14}$-$10^{-13}$ sec~\cite{PeltonOptExp2010}) compared to quantum emitters (QE, $\tau_{\rm QE}=10^{-9}$-$10^{-8}$ sec~\cite{PaspalakisJPhys2012}), experiments show that they are capable of handling quantum entanglement and nonclassical states for times longer than $\tau_{\rm ent}=10^{-10}$ sec~\cite{PlasmonEntanglementLifetime_NanoLett_2012,SqueezedPlasmons_PRL_2009,altewischer2002plasmon}. Entangled and nonclassical states, once observable in the far-field-coupled photons, are now producible in the near-field electromagnetic radiation, e.g. in the form of plasmon oscillations~\cite{SqueezedPlasmons_PRL_2009}. Fano resonances can also enhance the degree of entanglement~\cite{EntanglementFanoOptLett2012}.

It is well-demonstrated that presence of a metal nanoparticle~(MNP) near a quantum emitter~(QE) modifies (increases) the bandwidth of the QE about 3-4 orders of magnitude~\cite{FluorescenceEnhancementNanophotonics2015}, i.e. the Purcell effect. Even though radiation bandwidth of a bare QE, or a standard laser, is very narrow compared to the optical radiation; a QE coupled with a MNP, spaser (surface plasmon amplification by stimulated emission of radiation) nano-lasers~\cite{Spasers_Review2017,SpaserNanoprope2018}, radiate/lase in a very broad bandwidth~\cite{noginov2009demonstration}. This bandwidth modification enables the fast-turn on/off nano-dimensional lasers~(spasers), on one hand, enables miniaturized ultrafast-response~\cite{stockman2010spaser} technologies. On the other hand, they introduce a problem in the definition and quantification of entanglement/nonclassicality in such radiators.

Quantum entanglement witnesses and measures, we usually deal in quantum optics, rely on the inseparability of the two modes which are commonly represented by a single wave-vector ${\bf k}$, i.e. $\hat{a}_{{\bf k}_1}$ and  $\hat{a}_{{\bf k}_2}$. Here, $\omega_{1,2}=ck_{1,2}$ are the carrier frequencies of the two nonclassical beams. Such a treatment is acceptable for narrow-frequency-width pulses, especially when the detector is placed (measurement is performed) in the far-field, where choice of single component $k$ is justified also with the directional (small solid angle) arguments. Such a simplification, two single $k$ modes, can be applied also to the modified (very-broadened) emission of a QE-MNP hybrid for the far-field detection. Because a specific $k$ value is detected, again, due to the small solid angle argument. However, quantification of the entanglement/nonclassicality via detecting the inseparability of only the two modes, e.g. carrier frequencies of the two beams, is highly insufficient in the detection and "use" of the whole entanglement potential of the two pulses. Maximum entanglement/nonclassicality harvesting, e.g. in quantum teleportation~\cite{BraunsteinNatureTeleportation2015} and quantum thermodynamics~(heat engines)~\cite{MustecapQHeatEngine2016,MustecapQHeatEngine2018}, is important in the efficiency of such devices. The situation~(insufficiency) becomes even more adverse, if the quantification is tried to perform via two near-field detectors~\cite{PlasmonDetectionNanoLett2015}, where pronunciation of two modes becomes impossible.

Therefore, entanglement of two wavepackets, once could be questioned due to curiosity, now, became a necessity~\cite{spaser_Entanglement_PRB_2018} with the development of fast-response nano-control~\cite{SpaserNanoprope2018} and nano-imaging techniques~\cite{PlasmonDetectionNanoLett2015}. In this paper, we aim to extend the notion of, i.e., (i) two-mode entanglement~(TME) to the entanglement of two wavepackets~(WPs) each containing a broadband of frequency components. (ii) We also introduce a notion for the nonclassicality~(Nc) of a WP, which is referred as single-mode nonclassicality~(SMNc), e.g. squeezing, for an almost single-mode beam. Furthermore, (iii) we extend the definition of entanglement between an ensemble of QEs and the emitted-mode~\cite{PolzikNature2001ensemble} to the ensemble-WP entanglement~\cite{ZubairyQDs_Plasmon_entanglement}. 

After a survey among the possible extensions/generalizations of the entanglement into WPs, we demonstrate that the most meaningful definition could be performed via making a replacement, $\hat{a}\to\sum_{\bf r} \hat{a}_{\bf r}$, from a single-mode to a WP. The summation $\sum_{\bf r}$ stands for the volume/area of the detector for the measurement via a near-field detector and $\sum_{\bf r}$ stands for the whole space for the calculation of the total entanglement existing between the two WPs. $\hat{a}_{\bf r}$ is the operator annihilating a photon~(could as well be a plasmon) at position $\bf r$. In particular, we study the entanglement of WPs emitted either from two initially-entangled cavities or initially-entangled atoms.

The paper is organized as follows. First, in Sec.~\ref{sec:correlations}, we introduce the entanglement of two WPs using the electric fields of the two WPs, i.e. $\hat{a}_i \to \hat{E}_i^{(+)}=\sum_{\bf k} \varepsilon_k e^{i{\bf k}\cdot{\bf r}} \hat{a}_{\bf k}$. We show that generation (onset) of entanglement between the two pulses, at positions ${\bf r}_1$ and ${\bf r}_2$, propagates with the speed of light, $c$. This definition is demonstrated to be not useful for two purposes. ({\it 1}) Entanglement does not quantify the inseparability of the two WPs, but it witnesses on the inseparability (correlations) of the electric field measurements at the positions ${\bf r}_1$ and ${\bf r}_2$. ({\it 2}) Using such a definition, we face with a divergence problem, in $\sum{\bf k}\varepsilon_k$, when we desire to use the analogues of the stronger criterion Simon-Peres-Horedecki~(SPH)~\cite{SimonPRL2000} or the criterion by Duan-Giedke-Cirac-Zoller~(DGCZ)~\cite{DGCZ_PRL2000} for the WPs. We face the same divergence problem when we introduce $\hat{a}_i\to \sum_{{\bf k}_i} \hat{a}_{i,{\bf k}_i}$, although this definition has the potential to detect the inseparability of any two modes selected from each WPs. Next, in Sec.~\ref{sec:spatial}, we realize that, by introducing $\hat{a}_i \to \sum_{{\bf r}_i}\hat{a}_{i,{\bf r}_i}$, we can both circumvent the divergence problem in item ({\it 2}) and calculate the total entanglement which two near-field detectors measure. We can also calculate the whole entanglement (potential) between the two WPs. Here, $i=1,2$ refers to the two WPs. 

In Sec.~\ref{sec:WP-WP}, we define the total entanglement between two WPs by introducing the annihilation operator $\hat{A}_i=\sum_{{\bf r}_i}\hat{a}_{i,{\bf r}_i}$. We introduce the analogues of SPH~\cite{SimonPRL2000} and Hillery{\&}Zubairy~(HZ)~\cite{Hillery&ZubairyPRL2006}, also derived by Shchukin{\&}Vogel priorly~\cite{ShchukinVogelPRL2005}, criteria for WP-WP entanglement. We study the time development of the total entanglement of two WPs, emitted from two initially entangled cavities/atoms; using both HZ and SPH criteria. In Sec.~\ref{sec:ensemble-WP}, we introduce ensemble-WP entanglement criteria by replacing $\hat{a}_i\to\hat{A}_i$. We study the spontaneous emission of a single atom and superradiant single-photon emission from a many-particle  entangled ensemble. In Sec.~\ref{sec:WPNc}, we define the nonclassiality~(Nc) of a WP both via noise matrix of $\hat{X}$, $\hat{P}$ operators defined over $\hat{A}$ and via a beam-splitter~(BS): by measuring the WP-WP entanglement this nonclassical WP generates at the BS output. We show that (a) when some of the constituent modes of the WP are squeezed or (b) when two modes of the WP are entangled, WP becomes nonclassical, i.e. with reduced noise in a $\hat{X}_\phi$ operator, with $\hat{A}_\phi=e^{i\phi}\hat{A}$. Section~\ref{sec:summary} contains our summary.

\section{Correlations of Electric-field measurements} \label{sec:correlations}

Arriving a convenient definition, or a notion, for the entanglement of two wavepackets~(WPs) necessitates the exploration of the correlations between the electric~(E) fields of he two WPs at different positions ${\bf r}_1$ and ${\bf r}_2$. It is straight forward to see that one can obtain the same forms with the two criteria, DGCZ~\cite{SimonPRL2000} and HZ~\cite{Hillery&ZubairyPRL2006,ShchukinVogelPRL2005}, for $\hat{a}_1 \to \hat{E}^{(+)}_1({\bf r}_1)$ and $\hat{a}_2 \to \hat{E}^{(+)}_2({\bf r}_2)$ where
\begin{equation}
\hat{E}^{(+)}_i({\bf r}_i) = \sum_{{\bf k}_i} \varepsilon_{k_i} e^{i{\bf k}_i\cdot {\bf r}_i} \hat{a}_{i,{\bf k}_i}
\end{equation}
are the positive part of the electric field operators associated with the two WPs, $i=1,2$. Each WP has the momentum components $\hat{a}_{i,{\bf k}_i}$. $\varepsilon_{k_i}=\sqrt{\hbar ck_i/\epsilon_0 V_i}$ is the electric field of a single photon, depending on the quantization volume $V_i$ of the $i$th WP. Following the same steps given in Ref.~\cite{Hillery&ZubairyPRL2006}, the analogous form of the HZ criterion can be written as
\begin{eqnarray}
\lambda_{\rm HZ}=\langle \hat{E}^{(+)}_2({\bf r}_2) \hat{E}^{(-)}_2({\bf r}_2) \hat{E}^{(+)}_1({\bf r}_1) \hat{E}^{(-)}_1({\bf r}_1) \rangle \nonumber
\\
-|\langle \hat{E}^{(+)}_2({\bf r}_2) \hat{E}^{(-)}_1({\bf r}_1) \rangle|^2,
\end{eqnarray}
where $\lambda_{\rm HZ}<0$ witnesses the inseparability of the two WPs, or the presence of nonlocal correlations between E-field measurements of the two WPs at positions ${\bf r}_1$ and ${\bf r}_2$. $\hat{E}^{(-)}_i({\bf r}_i)$ is the hermitian conjugate of $\hat{E}^{(+)}_i({\bf r}_i)$. HZ criterion, also derived by Shchukin{\&}Vogel priorly~\cite{ShchukinVogelPRL2005} priorly in another context, does not lead to any divergence problem since it does not necessitate the evaluation of a term like $\langle\hat{E}^{(+)}_i({\bf r}_i) \hat{E}^{(-)}_i({\bf r}_i)\rangle$, in difference to SPH or DGCZ criteria. 

One can also derive DGCZ criterion for the entanglement of two WPs with the replacement $\hat{x}_1 \to \hat{E}_1({\bf r}_1)$ and $\hat{x}_2 \to \hat{E}_2({\bf r}_2)$ using the same arguments in Ref.~\cite{DGCZ_PRL2000}, i.e. Cauchy-Schwarz inequality for separable states. Here $\hat{E}_i({\bf r}_i)=\hat{E}^{(+)}_i({\bf r}_i)+\hat{E}^{(-)}_i({\bf r}_i)$ is the electric field operator. This criterion, however, is not a useful one since it contains terms like $\langle\hat{E}^{(+)}_i({\bf r}_i) \hat{E}^{(-)}_i({\bf r}_i)\rangle$ which do diverge. SPH criterion also includes similar divergent terms and does not have any practical use here.

Our experience shows us that DGCZ criterion works good for quadrature-squeezed like states, while HZ criterion works good mainly for number-squeezed like states and superpositions of Fock states~\cite{NhaPRA2006Fock_states}. Here, in this section, we consider the entanglement of two WPs, emitted from two initially entangled cavities, $|\psi(0)\rangle=a_1(0)|1\rangle_{\rm c_1}|0\rangle_{\rm c_2} + a_2(0)|0\rangle_{\rm c_1}|1\rangle_{\rm c_2} $ into two different reservoirs, or from two initially entangled atoms $|\psi(0)\rangle=a_1(0)|e\rangle|g\rangle+a_2(0)|g\rangle_1|e\rangle_2$. (We study the extended version of the system in Ref.~\cite{SunndenBirthofEntang_PRL_2008} where the reservoirs are treated as two single modes.) Fortunately, we can study the correlations in such a system. Because the system emits the superpositions of Fock states, where HZ criterion, do not diverge, can be used.

\begin{figure}
\includegraphics [width=0.5\textwidth]{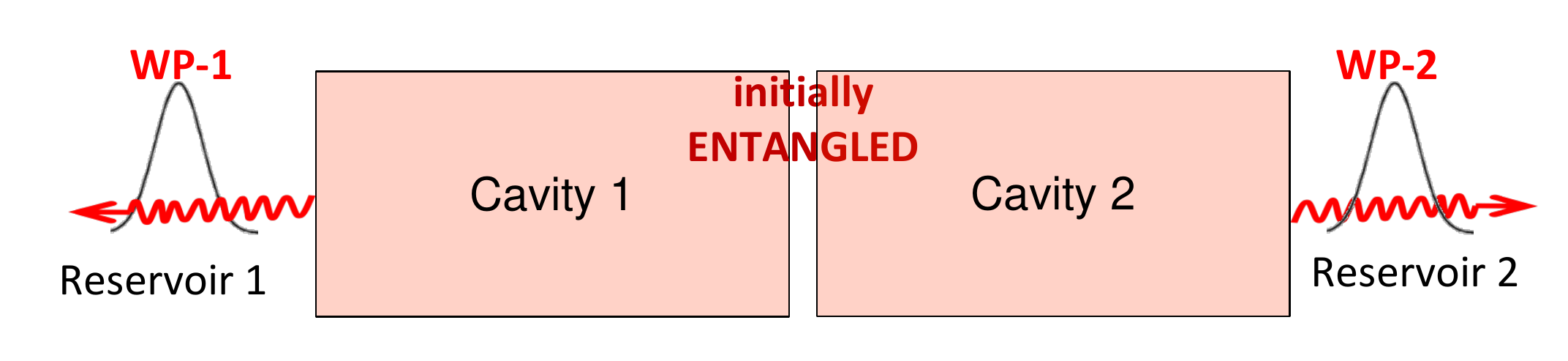} 
\caption{The two cavities are initially in an entangled state and they decay into two different reservoirs. We examine the time evolution of the onset of the entanglement of the two reservoirs, or in other words, correlations in the electric field measurements of the emitted wavepackets~(WPs) in the two reservoirs. We also calculate the total entanglement of the two WPs in Sec.~\ref{sec:WP-WP}.}
\label{fig1}
\end{figure}

In Fig.~\ref{fig1}, the two cavities are initially in an entangled state,  $|\psi(0)\rangle=\left( a_1(0)|1\rangle_{\rm c_1}|0\rangle_{\rm c_2} + a_2(0)|0\rangle_{\rm c_1}|1\rangle_{\rm c_2}\right) |0\rangle_{\rm R_1}|0\rangle_{\rm R_2}$, where $|\:\rangle_{\rm c_{1,2}}$ and $|\:\rangle_{\rm R_{1,2}}$ are the Fock states for the two entangled cavities and the two reservoirs the cavities decay, respectively. The solution of the interaction picture hamiltonian~\cite{SunndenBirthofEntang_PRL_2008}
\begin{equation}
\hat{V}=\sum_{i=1}^2 \sum_{{\bf k}_i} \hbar g_{{\bf k}_i} \hat{a}_{i,{\bf k}_i}^\dagger \hat{c}_i 
\: e^{-i(\Omega_i-\omega_{{\bf k}_i})t} \; + \; H.c.
\end{equation}
in subspace of possible states
\begin{eqnarray}
|\psi(t)\rangle=&&\left( b_1(t)|0\rangle_{\rm c_1} |1\rangle_{\rm c_2} + b_2(t)|1\rangle_{\rm c_1} |0\rangle_{\rm c_2} \right) |0\rangle_{\rm R_1} |0\rangle_{\rm R_2} \qquad
\nonumber
\\
+&& |0\rangle_{\rm c_1} |0\rangle_{\rm c_2}   \Big(  \sum_{{\bf k}_1} d_{1,{{\bf k}_1}}(t) |1_{{\bf k}_1}\rangle_{\rm R_1} |0\rangle_{\rm R_2}  \nonumber 
\\ 
&& \hspace{1.4 cm}+ |0\rangle_{\rm R_1} \sum_{{\bf k}_2} d_{2,{{\bf k}_2}}(t) |1_{{\bf k}_2}\rangle_{\rm R_2}  \Big)
\label{psitwocav}
\end{eqnarray}
is determined by the coefficients
\begin{eqnarray}
&&b_i(t)=e^{-\gamma_it/2} a_i(0) ,
\\
&&d_{i,{\bf k}_i}(t) = g_{k_i} a_i(0) \frac{1-e^{-i(\Omega_i-\omega_{k_i})t - \gamma_it/2}}{(\omega_{k_i}-\Omega_i)+i\gamma_i/2},
\label{cavitydi}
\end{eqnarray}
where $\Omega_{i}$ and $\gamma_i$ are the cavity resonance and damping rate, respectively. $g_{k_i}$ is the coupling strength between the $i$th cavity and the $i$th reservoir. When we consider sufficiently long two cavities, and thin mirrors which couple the cavities to the reservoirs, HZ criterion for the entanglement of the two WPs can be calculated as
\begin{eqnarray}
&&\lambda_{\rm HZ}(t)\simeq -(2\pi)^2 g_1^2(\Omega_1) D_1(\Omega_1) g_2^2(\Omega_1) D_2(\Omega_2) \varepsilon_{K_1} \varepsilon_{K_2}
\nonumber \\
\times && e^{-\gamma_1|z_1-ct|/2c} \: e^{-\gamma_2|z_2-ct|/2c} \: \Theta(t-z_1/c)\: \Theta(t-z_2/c), \qquad
\label{lamdaHZE}
\end{eqnarray}  
where we assume that dispersion of the cavity emission is negligible in the transverse directions, $\hat{x}_i$ and $\hat{y}_i$.  $D(\Omega_i)$ is the density of states at the cavity resonance $\Omega_i$ and can be related to the damping rate as $\gamma_i=\pi D_i(\Omega_i) g^2({\Omega_i})$. $\varepsilon_{K_1}=\sqrt{\hbar\Omega_i/\epsilon_0 V_i}$ with $K_i=\Omega_i/c$. The step functions in Eq.~(\ref{lamdaHZE}), $\Theta(t-z_i/c)$ reveal the luminal "onset" of correlations (entanglement) between the two WPs, at $z_1$ and $z_2$. We note that this approximate result for entanglement is realistic in the following aspect. For two collimated wavepackets of narrow frequency band, the entanglement does not decay (or decays negligibly) with $z$-propagation. We also evaluate the $\lambda_{\rm HZ}(t)$ for an uncollimated emission, where we find that absolute value of its negativity decreases with spatial spreading.

Such a definition of entanglement (correlations) between two WPs is instructive especially for exploring the onset of the entanglement in spatial dimensions. However, such a definition fails to work for most useful nonclassical states, the Gaussian states, which are the ones convenient to generate and use in the experiments.

Moreover, it has a potential only to quantify the WP-WP entanglement on a position-to-position basis. That is, it does not quantify the "total" entanglement between the two WPs. A candidate for quantifying the total entanglement, i.e. between all of the modes, could be
\begin{equation}
\hat{a}_i \to \sum_{ {\bf k}_i} \hat{a}_{i,{\bf k}_i} \quad {\rm or} \quad \hat{a}_i \to \sum_{ {\bf k}_i} \varepsilon_{k_i} \hat{a}_{i,{\bf k}_i},
\end{equation} 
which has the potential to address the entanglement of any two modes, $\hat{a}_{1,{\bf k}_1}$ and $\hat{a}_{2,{\bf k}_2}$, between the two WPs~
\footnote{We use the phrase "has the potential to detect entanglement" on purpose. Because noise reduction due to $\hat{a}_{1,{\bf k}_1} \leftrightarrow \hat{a}_{2,{\bf k}_2}$ entanglement can be screened by a noise increase due to two other modes $\hat{a}_{1,{\bf k}'_1} \leftrightarrow \hat{a}_{2,{\bf k}'_2}$. \label{potential}}. Such a definition however is again not useful for Gaussian states since it leads to divergence in SPH and DGCZ criteria.

\section{Convenience of working in the spatial domain ---convergence} \label{sec:spatial}

Next, we realize that we cannot avoid the divergence of $\sum_{{\bf k}}$ summation, since we cannot adopt a bound for the ${\bf k}$-space. In difference to momentum space, fortunately, a $\sum_{{\bf r}}$ summation is bound by the volume $V$ which can be handled theoretically or can be limited in the experiments. Thus, we choose to work in the spatial domain by introducing the mode expansion~\cite{MeystrePRL1999_BEC_SR,tasginPRA2011vortex}
\begin{equation}
\hat{a}({\bf r})=\sum_{\bf k} e^{i{\bf k}\cdot{\bf r}} \hat{a}_{\bf k},
\label{ar}
\end{equation}
which can be Fourier transformed as 
\begin{equation}
\sum_{\bf r} \hat{a}({\bf r})   e^{-i{\bf k}\cdot{\bf r}} = \sum_{{\bf k}'} \left( \sum_{\bf r} e^{i({\bf k}-{\bf k}')\cdot{\bf r}}    \right) \hat{a}_{{\bf k}'}= \hat{a}_{\bf k}
\label{ak}
\end{equation}
by defining the normalized summation $\sum_{\bf r} \to \int d^3{\bf r}/V$ and using $\sum_{\bf k} \to \frac{V}{(2\pi)^3} \int d^3{\bf k}$ as usual~\cite{ScullyZubairyBook}.  Hermitian conjugates of Eqs.~(\ref{ar}) and (\ref{ak}) can be used, applied on vacuum, to relate the spatial and momentum Fock spaces, e.g., as
\begin{equation}
|1_{\bf r}\rangle=\sum_{\bf k} e^{-i{\bf k}\cdot{\bf r}}|1\rangle_{\bf k} \quad {\rm and} \quad
|1_{\bf k}\rangle=\sum_{\bf k} e^{i{\bf k}\cdot{\bf r}}|1_{\bf r}\rangle.
\end{equation}

The advantage of working in the spatial domain, by defining the annihilation operator
\begin{equation}
\hat{a}_i \to \hat{A}_i = \sum_{{\bf r}_i} \hat{a}_i({\bf r}_i)
\label{A}
\end{equation}
is, now, the quantity $\langle\hat{A}_i\hat{A}_i^\dagger\rangle$ does not diverge! Here, $i=1,2$ enumerates the two WPs. Moreover, Eq.~(\ref{A}), when used in an entanglement criterion, has the potential~${}^{\ref{potential}}$ to detect correlations between any two spatial modes, $\hat{a}_{1,{\bf r}_1} \leftrightarrow \hat{a}_{2,{\bf r}_2}$, of the two WPs. One can obtain the commutation
\begin{equation}
[\hat{A},\hat{A}^\dagger]=1
\label{commA}
\end{equation}
from the relation $[\hat{a}({\bf r}),\hat{a}({\bf r}')]=V\delta({\bf r}-{\bf r}')$ which deduces from Eq.~(\ref{ar}) and $[\hat{a}_{\bf k},\hat{a}_{{\bf k}'}]=\delta_{ {\bf k},{\bf k}' }$. Commutation~(\ref{commA}) remains convergent and dimensionless via normalized definition of the spatial integration $\sum_{\bf r} \to \frac{1}{V} \int d^3{\bf r}$.

In the next section, we use the annihilation operator $\hat{A}$, defined in Eq.~(\ref{A}), to obtain WP analogues of DGCZ~\cite{DGCZ_PRL2000}, HZ~\cite{Hillery&ZubairyPRL2006} and SPH~\cite{SimonPRL2000} criteria. We also use the same form, $\hat{A}$, for introducing the ensemble-WP entanglement (Sec.~\ref{sec:ensemble-WP}) and nonclassicality of a WP, in Sec.~\ref{sec:WPNc}).

\section{Wavepacket--Wavepacket entanglement} \label{sec:WP-WP}

In order to obtain a "convergent" entanglement criterion which has the potential~${}^{\ref{potential}}$ to address a kind of "total" entanglement, e.g. taking all spatial or $k$-mode correlations into account, we introduce $\hat{A}_i=\sum_{{\rm r}_i} \hat{a}_i({\rm r}_i)$, for instance, for the DGCZ criterion~\cite{DGCZ_PRL2000} 
\begin{equation}
\lambda_{\scriptscriptstyle{\rm DGCZ}}= \langle(\Delta\hat{u})^2\rangle + \langle(\Delta\hat{v})^2\rangle-(\alpha^2+\beta^2),
\end{equation}
where $\lambda_{\scriptscriptstyle{\rm DGCZ}}<0$ witnesses the inseparability of the two WPs. Here, the operators are
\begin{eqnarray}
\hat{u}=\alpha \hat{X}_1 + \beta \hat{X}_2,
\\
\hat{v}=\alpha \hat{P}_1 - \beta \hat{P}_2,
\end{eqnarray}
where
\begin{eqnarray}
\hat{X}_i = (\hat{A}_i^\dagger + \hat{A}_i)/\sqrt{2}=\sum_{{\bf r}_i } \hat{x}_i({\bf r}_i),
\label{Xi}
\\
\hat{P}_i = i(\hat{A}_i^\dagger - \hat{A}_i)/\sqrt{2}=\sum_{{\bf r}_i } \hat{p}_i({\bf r}_i).
\label{Pi}
\end{eqnarray}
$\hat{X}_i$ and $\hat{P}_i$ satisfy the usual commutation relation
\begin{equation}
[\hat{X}_i,\hat{P}_i]=i.
\label{commXP}
\end{equation}

Eq.~(\ref{commXP}) is a central result of the paper. Because it indicates that any two-mode entanglement~(TME) criterion derived for $\hat{a}_1 \leftrightarrow \hat{a}_2$, see also Ref.~\cite{tasgin2019anatomy}, are valid also for the inseparability of the two WPs, when $\hat{X}_i$ and $\hat{P}_i$ are defined as in Eqs.~(\ref{Xi}) and (\ref{Pi}).

More explicitly, if one defines the operators
\begin{equation}
\hat{\xi}=[\hat{X}_1 \; \hat{P}_1 \; \hat{X}_2 \; \hat{P}_1 ]
\end{equation}
and calculates the noise matrix
\begin{equation}
V_{ij}=\frac{1}{2}\langle \hat{\xi}_i \hat{\xi}_j + \hat{\xi}_j \hat{\xi}_i \rangle = \langle\hat{\xi}_i\rangle \langle\hat{\xi}_j\rangle,
\end{equation}
the SPH criterion~\cite{SimonPRL2000}
\begin{eqnarray}
\lambda_{\scriptscriptstyle{\rm SPH}}=&& \det{A} \det{B}  + \left(\frac{1}{4} - |\det C| \right)^2 - {\rm tr}(AJCJBJC^TJ) \nonumber
\\
-&&\frac{1}{4}(\det A+\det B)
\label{lambdaSPH}
\end{eqnarray}
is also valid for the entanglement of two WPs. Here, $A$,$B$ and $C$ are 2$\times$2 matrices defining the 4$\times$4 noise matrix $V=[A\:,\:C\:;\:C^T\:,\:B]$. SPH criterion~\cite{SimonPRL2000} is a particularly important one, since it accounts any intra-mode rotations, i.e. $\hat{A}_\phi=e^{i\phi}\hat{A}$, in the $X_i$-$P_i$ plane~\cite{tasgin2019anatomy}.

In  Sec.~III.3 of Ref.~\cite{tasgin2019anatomy}, we show that such a strong criterion is possible to be derived also for number-phase squeezed like states~\cite{PeggJModOpt1990Intelligent}. Similar to SPH criterion~\cite{SimonPRL2000}, it accounts intra-mode rotations in the $n$-$\Phi$, number-phase, plane. This new criterion is also valid for detecting the entanglement of two WPs. 

Similarly, Hillery{\&}Zubairy~(HZ) criterion~\cite{Hillery&ZubairyPRL2006}, also formerly introduced by Shchukin and Vogel~\cite{ShchukinVogelPRL2005}, 
\begin{equation}
\lambda_{\scriptscriptstyle{\rm HZ}} = \langle \hat{A}_2^\dagger\hat{A}_2\hat{A}_1^\dagger\hat{A}_1 \rangle
-|\langle\hat{A}_2^\dagger \hat{A}_1\rangle|^2
\label{lambdaHZ}
\end{equation}
can be derived, using the same arguments in Ref.~\cite{Hillery&ZubairyPRL2006}, for two WPs.
\subsection{Two entangled cavities}
In the following, we calculate the total entanglement between two wavepackets~(WPs) emitted from two initially entangled cavities into two different reservoirs. This is depicted in Fig.~\ref{fig1}. First, we calculate the $\lambda_{\scriptscriptstyle{\rm HZ}}(t)$ given in Eq.~(\ref{lambdaHZ}), since the emitted pulses are superpositions of Fock states. Second, we preform the same calculation for $\lambda_{\scriptscriptstyle{\rm SPH}}$ given in Eq.~(\ref{lambdaSPH}). Similar results can be obtained also for the emission of two initially entangled atoms.
\subsubsection{HZ criterion:}
The solution of the emission from two entangled cavities, Eq.~(\ref{psitwocav}), can be transformed to the spatial domain of the two reservoirs as
\begin{eqnarray}
|\psi(t)\rangle=&&\left( b_1(t)|0\rangle_{\rm c_1} |1\rangle_{\rm c_2} + b_2(t)|1\rangle_{\rm c_1} |0\rangle_{\rm c_2} \right) |0\rangle_{\rm R_1} |0\rangle_{\rm R_2} \qquad
\nonumber
\\
+&& |0\rangle_{\rm c_1} |0\rangle_{\rm c_2}   \Bigg[  |0\rangle_{\rm R_1} \Big(\sum_{{\bf r}_2} I_2({\bf r}_2,t) |1_{{\bf r}_2}\rangle_{\rm R_2} \Big)
\\
&&\hspace{1.4 cm} +\Big(\sum_{{\bf r}_1} I_1({\bf r}_1,t) |1_{{\bf r}_1}\rangle_{\rm R_1} \Big) |0\rangle_{\rm R_2} \Bigg],
\label{psitwocavR}
\end{eqnarray}
where $I_i({\bf r}_i,t)=\sum_{{\bf k}_i}d_{i,{\bf k}_i}(t) e^{i{\bf k}_i\cdot{\bf r}_i }$ with $d_{i,{\bf k}_i}(t)$ is given in Eq.~(\ref{cavitydi}). Using the contour-integration method, momentum integral can be calculated as
\begin{eqnarray}
I_i({\bf r}_i,t)=\frac{Vb_i(0)}{2\pi cr_i}K_i g_i(\Omega_i) e^{-(i\Omega_i+\gamma_i/2)r_i/c} \Theta(ct-r_i),\hspace{0.9 cm}
\end{eqnarray}
where $K_i=\Omega_i/c$ and $g_i(\Omega_i)$ is the cavity-reservoir coupling evaluated at the cavity resonance $\omega=\Omega_i$.
We remark that, in the evaluation of $I_i$ we did not make a collimated-beam approximation, i.e. ${\bf k}\simeq k_z$, which we performed in Eq.~(\ref{lamdaHZE}). In Eq.~(\ref{lamdaHZE}), we perform collimated-beam approximation for providing an easier understanding on the experiments. The notion of entanglement would not change if we were/were not performed such an approximation.

\begin{figure}
\includegraphics [width=0.47\textwidth]{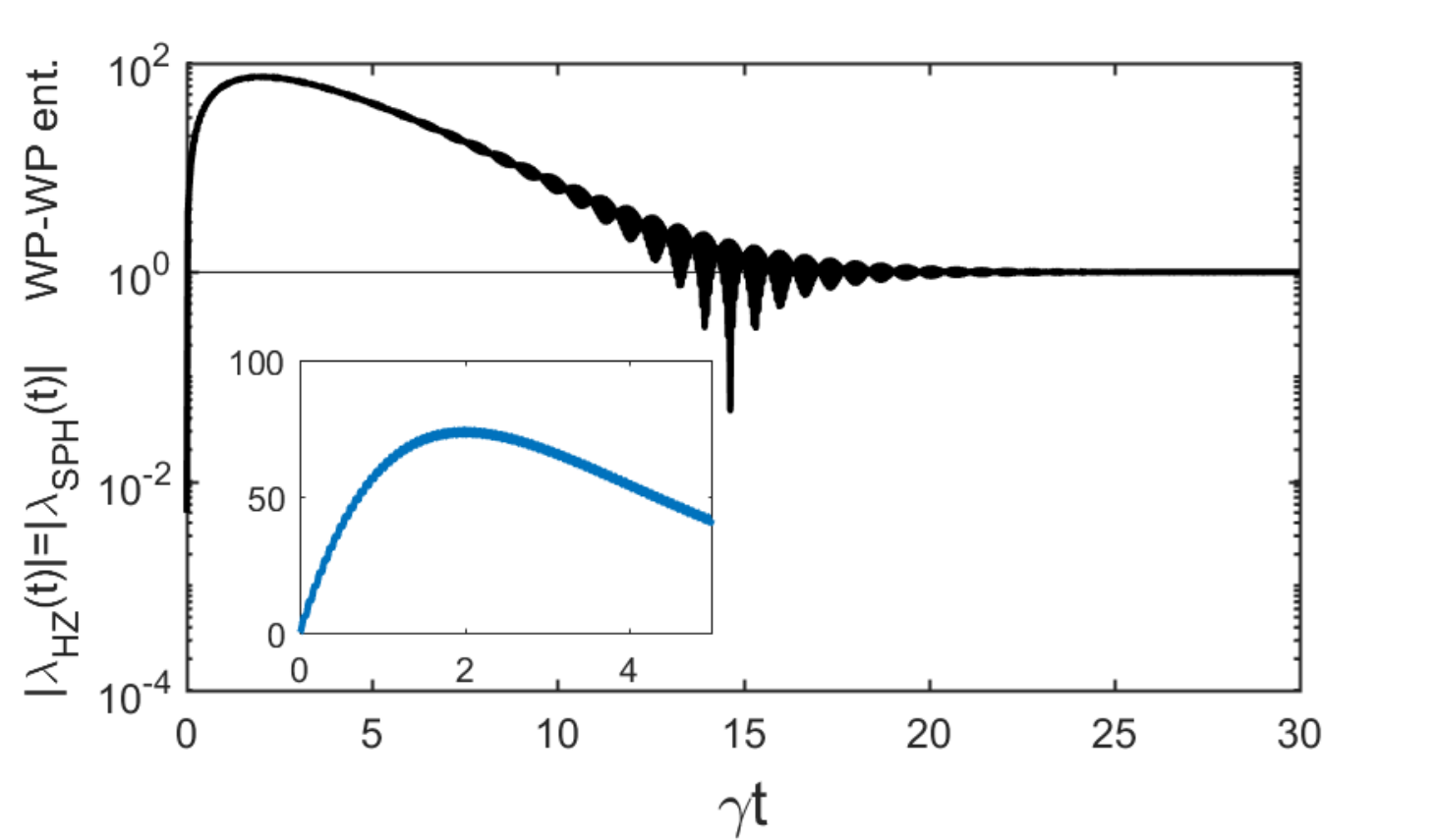} 
\caption{Hillery{\&}Zubairy and Simon-Peres-Horodecki criteria, $\lambda_{\scriptscriptstyle{\rm HZ}}(t)=\lambda_{\scriptscriptstyle{\rm SPH}}(t)$, for the two wavepackets emitted from two initially entangled cavities of Fig.~\ref{fig1}. In difference to point-wise, $E_1({\bf r}_1,t)\leftrightarrow E_2({\bf r}_2,t)$, E-field correlations studied in Sec.~\ref{sec:correlations}, $\lambda_{\scriptscriptstyle{\rm HZ, SPH}}(t)<0$ witnesses a kind of total entanglement between the two WPs emitted into two different reservoirs.}
\label{fig2}
\end{figure}

When $\hat{A}_1$ operator is acted on the $|\psi(t)\rangle$, we obtain
\begin{eqnarray}
\hat{A}_1|\psi(t)\rangle = &&\Big(  \sum_{{\bf r}_1} \sum_{{\bf r}_1'} I_1({\bf r}_1,t) \hat{a}_1({\bf r}_1') |1_{{\bf r}_1}\rangle_{\rm R_1} \Big)
 |0\rangle_{\rm R_2} |0\rangle_{\rm c_1}|0\rangle_{\rm c_2}
 \nonumber \\
=&&\Big(  \sum_{{\bf r}_1}  I_1({\bf r}_1,t)  \Big)
  |0\rangle_{\rm R_1} |0\rangle_{\rm R_2} |0\rangle_{\rm c_1}|0\rangle_{\rm c_2}.
  \label{A1psit}
\end{eqnarray}
The same form appears for $(\hat{A}_2|\psi(t)\rangle)^\dagger=\langle\psi(t)|\hat{A}_2^\dagger$. If we define the spatial integral in Eq.~(\ref{A1psit}) as $J_i(t)= \sum_{{\bf r}_i}  I_i({\bf r}_i,t)$, the second term of the $\lambda_{\scriptscriptstyle{\rm HZ}}$, in Eq.~(\ref{lambdaHZ}) can be identified as $|\langle\hat{A}_2^\dagger\hat{A}_1\rangle|^2$. It is evident from Eq.~(\ref{A1psit}) is that $\hat{A}_2\hat{A}_1|\psi(t)\rangle=0$. Hence, the first term in Eq.~(\ref{lambdaHZ}) is zero. Then, HZ criterion for two WPs reduces to
\begin{equation}
\lambda_{\scriptscriptstyle{\rm HZ}}(t)=-|J_1(t)|^2\:|J_2(t)|^2,
\end{equation}
where spatial integrals can be evaluated as
\begin{eqnarray}
J_i(t)=\frac{2b_i(0)}{c}K_ig_i(\Omega_i) \frac{1-e^{\alpha_ict}+e^{\alpha_ict}\alpha_i ct}{\alpha_i^2},
\label{Ji}
\end{eqnarray}
with $\alpha_i ct=-(i\Omega_i+\gamma_i/2)t$, which do not depend on the reservoir volume. In Fig.~\ref{fig2}, we plot $\lambda_{\scriptscriptstyle{\rm HZ}}(t)$.  The total entanglement increases till the two WPs leave the two cavities (or the two atoms) completely. Then, it drops but approaches a constant value as $\gamma t \gg 1$. We scale the $y$-axis of Fig.~\ref{fig2} with $4a(0)b(0)K_1K_2g_1(\Omega_1)g_2(\Omega_2)/ c^2\alpha_1^2 \alpha_2^2$. We consider emission from a plasmonic cavity, thus choose $\gamma=10^{-2}\Omega$ with $\Omega$ is in the optical regime.

\subsubsection{SPH criterion:}
We can also calculate the total entanglement between the two WPs, using the SPH criterion defined in Eq.~(\ref{lambdaSPH}). The terms like $\langle\hat{A}_i^2\rangle$ and $\langle\hat{A}_2\hat{A}_1\rangle$ do vanish. So, the 2$\times$2 matrices become
\begin{eqnarray}
A=\begin{bmatrix} 
\ell_1 & 0
\\
0 & \ell_1
\end{bmatrix},\;
B=\begin{bmatrix} 
\ell_2 & 0
\\
0 & \ell_2
\end{bmatrix},\;{\rm and} \;
C=\begin{bmatrix} 
a & b
\\
-b & a
\end{bmatrix}, \hspace{0.5 cm}
\end{eqnarray}
where $\ell_{1,2}=\frac{1}{2}+|J_{1,2}|^2$, $a=(J_2^*J_1+J_1^*J_2)/2$ and $b=i(J_2J_1^*-J_1J_2^*)/2$. The SPH criterion is evaluated as
\begin{equation}
\lambda_{\scriptscriptstyle{\rm SPH}}=\ell_1^2\ell_2^2 + \Big( \frac{1}{4} - (a^2+b^2)^2\Big)^2 -2\ell_1\ell_2(a^2+b^2)-\frac{1}{4}(\ell_1^2+\ell_2^2),
\end{equation}
which reduces to
\begin{equation}
\lambda_{\scriptscriptstyle{\rm SPH}}(t)=-|J_1(t)|^2\: |J_2(t)|^2=\lambda_{\scriptscriptstyle{\rm HZ}}
\end{equation}
for the particular system we consider here.

\section{Ensemble-Wavepacket entanglement} \label{sec:ensemble-WP}

Similarly, we can introduce an entanglement criterion between an ensemble and a (e.g. emitted) WP. When we change $\hat{a}\to\hat{A}$ in the Eq.~(4) of Ref.~\cite{tasgin2017many}, it is straightforward to obtain the criterion
\begin{equation}
\mu_{\scriptscriptstyle{\rm HZ}}=\langle\hat{S}_+\hat{S}_-\hat{A}^\dagger\hat{A}\rangle - |\langle\hat{S}_+\hat{A}\rangle|^2,
\label{muHZ}
\end{equation} 
which works better for the entanglement of number (Fock) like states with an ensemble. This is the case for the spontaneous emission of a single atom~\cite{ScullyZubairyBook} or superradiant single-photon emission from an ensemble of many-particle entangled atoms~\cite{tasgin2017many,svidzinskyPRA2008cooperative,ScullyScience2009SR}. Here, $\hat{S}_+=\sum_{j=1}^N \sigma_j^{(+)}$ is the collective raising operator for the ensemble containing $N$ two-level atoms with $\sigma_j^{(+)}$ is the Pauli matrix of the $j$th atom, and $\hat{S}_-=\hat{S}_+^\dagger$.

One can also obtain the analogue of DGCZ criterion for ensemble-WP entanglement, $\hat{a}\to\hat{A}$ in Ref.~\cite{PolzikNature2001ensemble}, by examining the uncertainty bound for $\langle(\Delta\hat{u})^2\rangle + \langle(\Delta\hat{v})^2\rangle$ using
\begin{eqnarray}
\hat{u}=\hat{S}_x + \hat{X} \quad {\rm and} \quad \hat{v}=\hat{S}_y - \hat{P},
\end{eqnarray}
where $\hat{X}=(\hat{A}^\dagger+\hat{A})/\sqrt{2}$, $\hat{P}=i(\hat{A}^\dagger-\hat{A})/\sqrt{2}$, $\hat{S}_x=(\hat{S}_++\hat{S}_-)/2$ and $\hat{S}_y=i(\hat{S}_--\hat{S}_+)/2$. Such a criterion has already been studied for the entanglement between an ensemble and  a single mode of light~\cite{PolzikNature2001ensemble}, in the context of squeezing transfer from a nonclassical light to an ensemble resulting in spin squeezing. Here, we only make the replacement $\hat{a}\to\hat{A}$ and introduce ensemble-WP entanglement. DGCZ criterion works fine for Gaussian or quadrature-squeezed like states.

Below, first, we calculate the $\mu_{\scriptscriptstyle{\rm HZ}}(t)$ for the spontaneous emission of a single atom. Next, we evaluate $\mu_{\scriptscriptstyle{\rm HZ}}(t)$ for single-photon superradiant emission~\cite{ScullyScience2009SR,svidzinskyPRA2008cooperative} from an initially entangled ensemble of atoms~\cite{tasgin2017many}.

\subsection{Spontaneous emission of a single atom}

\begin{figure}
\includegraphics[width=0.47 \textwidth]{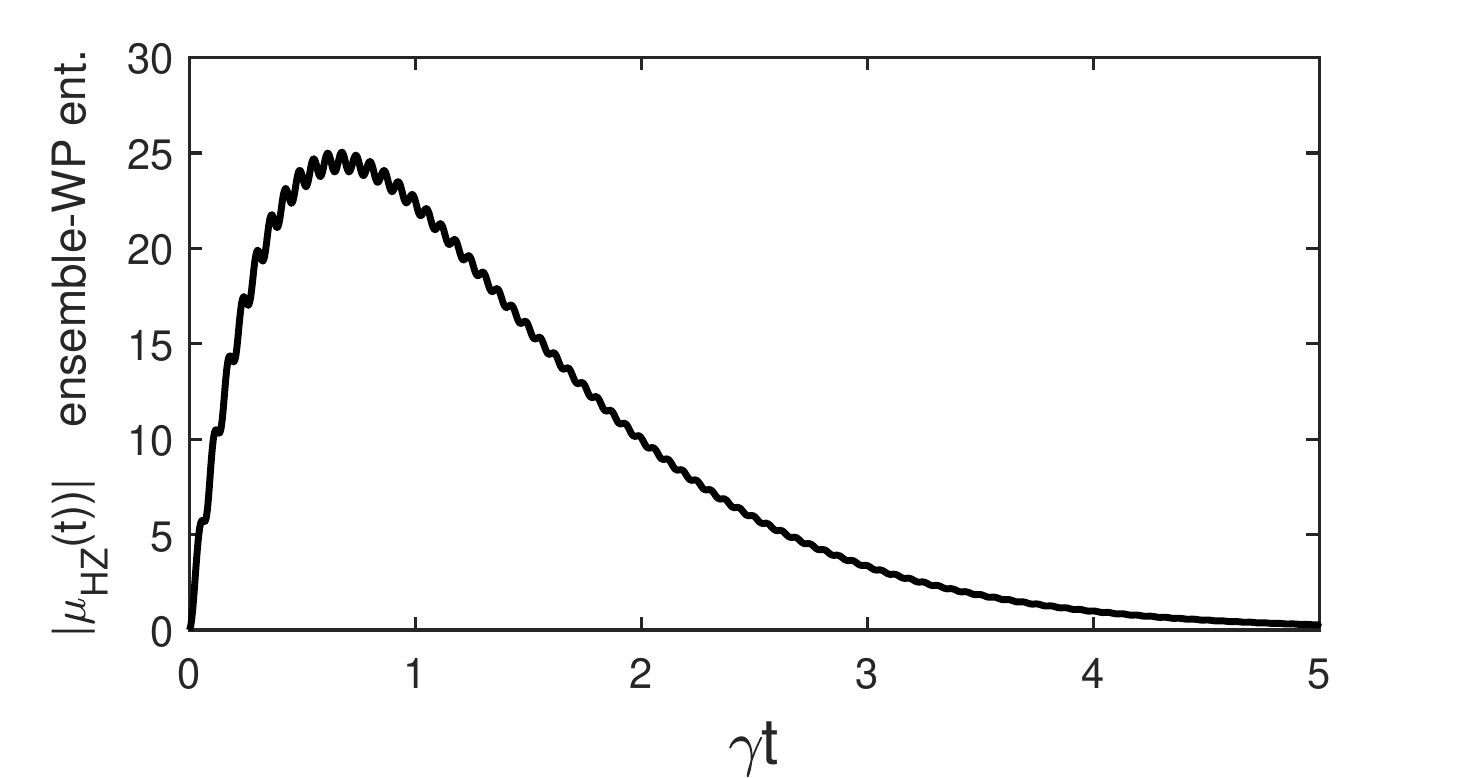}
\caption{Spontaneous emission from a single atom. Evolution of the entanglement  $\mu_{\scriptscriptstyle{\rm HZ}}(t)$<0 between the atom and the  emitted wavepacket.  Superradiant single-photon emission from an ensemble shows a similar behaviour except emission time determined by collective decay $\gamma_N$ in place of single atom decay $\gamma$.}
\label{fig3}
\end{figure}

The wave function of a two-level atom, initially in the excited state, is give by~\cite{ScullyZubairyBook}
\begin{equation}
|\psi(t)\rangle=\beta(t) |e\rangle|0\rangle + |g\rangle \sum_{{\bf k}} \gamma_{\bf k}(t) |1_{\bf k}\rangle,
\end{equation}
where spontaneous emission is possible into many $\bf k$ modes with probability amplitudes
\begin{equation}
\gamma_{\bf k}(t)= e^{-i{\bf k}\cdot{\bf r}_0} g_k \frac{1-e^{i(\omega_k-\omega_{eg})t-\Gamma t/2}}{(\omega_k-\omega_{eg})+i\Gamma/2},
\end{equation}
where ${\bf r}_0$ is the position of the atom and $\beta(t)=e^{-\Gamma t/2}$. $\omega_{eg}$ and $\Gamma$ are the level-spacing and damping rate of the atom, respectively. $g_k$ is the coupling strength of the ${\bf k}$ vacuum mode with the atomic dipole. When $\hat{A}$ acts on this state, it results 
\begin{equation}
\hat{A}|\psi(t)\rangle= \left[\sum_{\bf r} \left( \sum_{\bf k} e^{i{\bf k}\cdot{\bf r}} \gamma_{\bf k}(t) \right)  \right] |g\rangle | 0\rangle ,
\label{Apsit}
\end{equation}
where $\sum_{\bf k}$ integration in the inner parenthesis, $I_A$, yields 
\begin{equation}
I_A({\bf r},t)= \frac{V}{2\pi\: c r_s} g(\omega_{eg}) K_{eg} e^{-(i\omega_{eg}+\Gamma/2)r_s/c} \Theta(ct-r_s),
\label{IA_SE}
\end{equation}
with $r_s=|{\bf r}-{\bf r}_0|$, $K_{eg}=\omega_{eg}/c$ and $\Theta(x)$ is the step-function. Then, the $\sum_{\bf r}$ spatial integration results
\begin{equation}
J_A(t)=\frac{2g(\omega_{eg})K_{eg}}{c} \frac{1-e^{\alpha ct}+e^{\alpha ct}\alpha ct}{\alpha^2},
\label{JA_SE}
\end{equation}
similar to Eq.~(\ref{Ji}) of the previous section. Here, $\alpha c t=-(i\omega_{eg}+\Gamma/2)t$. It is easy to see from Eq.~(\ref{Apsit}) that $\hat{S}_-\hat{A}|\psi(t)\rangle=0$ which turns the first term in $\mu_{\scriptscriptstyle{\rm HZ}}$, Eq.~(\ref{muHZ}), equal to zero. The $(\langle\psi(t)|\hat{S}_+)^\dagger=\hat{S}_-|\psi(t)\rangle$ is
\begin{equation}
\hat{S}_-|\psi(t)\rangle=\beta(t)|g\rangle|0\rangle.
\end{equation}
So, HZ criterion becomes
\begin{equation}
\mu_{\scriptscriptstyle{\rm HZ}}(t) = -|\beta(t)|^2\:|J_A(t)|^2=-e^{-\Gamma t}\:|J_A(t)|^2.
\end{equation}
In Fig.~\ref{fig3}a, we plot $\mu_{\scriptscriptstyle{\rm HZ}}(t)$. 

\subsection{Superradiant emission from an ensemble}

We also study the entanglement of the superradiantly emitted single photon from an initially entangled ensemble of atoms $|\phi(0)\rangle_{\rm ens} = \sum_{j=1}^N e^{i{\bf k}_0\cdot {\bf r}_j} |e_j\rangle$, where $|e_j\rangle$ indicates that the $j$th atom is in the excited state and all other $(N-1)$ ones are in the ground state. The method for the generation of such a state is described in Ref.~\cite{ScullyScience2009SR}. ${\bf r}_j$ are the positions of the atoms in the ensemble which can be much larger than the emission wavelength $\lambda_0=2\pi/k_0$. In Fig.~(4) of Ref.~{\cite{tasgin2017many}, we demonstrated the entanglement between the central mode (carrier frequency) of the emitted light and the ensemble. Here, in difference, we examine the entanglement of the ensemble with the whole emitted light, the wavepacket~(WP).

Time evolution, superradiant emission, of this initial state is given~\cite{svidzinskyPRA2008cooperative} by
\begin{equation}
|\psi(t)\rangle = \sum_{j=1}^N \beta_j(t) |e_j\rangle|0\rangle + \Big( \sum_{{\bf k}} \gamma_{\bf k}(t) |1_{\bf k}\rangle  \Big) |g\rangle,
\end{equation}
where
\begin{eqnarray}
&&\beta_j(t)=\frac{1}{\sqrt{N}} e^{-\gamma_Nt} e^{i{\bf k}_0\cdot{\bf r}_j},
\\
&&\gamma_{\bf k}(t) = \frac{g_k}{\sqrt{N}} \frac{1-e^{-\gamma_Nt+i(\omega_k-\omega_{eg})t}}{(\omega_k-\omega_{eg}+i\gamma_N)}
\sum_{j=1}^N e^{i({\bf k}_0-{\bf k})\cdot {\bf r}_j}. \qquad
\end{eqnarray}
This emission, from an extended ($L>\lambda_0$) entangled ensemble, is referred as timed superradiance and the initial state is called as timed-Dicke states. Here, $\gamma_N$ is the collective (superradiant) decay rate, which can be much larger than the decay rate of a single atom~\cite{svidzinskyPRA2008cooperative}.

$\hat{A}|\psi(t)\rangle$ can be calculated similar to the spontaneous emission case, where now $I_A$ in Eq.~(\ref{IA_SE}) becomes
\begin{eqnarray}
I_A^{\rm (SR)}({\bf r},t)= \sum_{j=1}^N && \frac{e^{i{\bf k}_0\cdot{\bf r}_j}}{\sqrt{N}}
 \frac{V}{2\pi\: c r_j} g(\omega_{eg}) K_{eg}  \hspace{2 cm}
 \nonumber \\
 &&\times e^{-(i\omega{eg}+\gamma_N/2)r_j/c} \: \Theta(ct-r_j).
\end{eqnarray}
$J_A^{\rm (SR)}(t)=\sum_{\bf r} I_A^{\rm (SR)}$ can also be calculated similarly which results
\begin{equation}
J_A^{\rm (SR)}(t)=J_A(t,\gamma_N)\sum_{j=1}^N \frac{e^{i{\bf k}_0\cdot {\bf r}_j}}{\sqrt{N}},
\label{JASR}
\end{equation}
where $J_A(t,\gamma_N)$ is the integral calculated for a single atom emission  in Eq.~(\ref{JA_SE}), with $\Gamma/2\to\gamma_N$. We define the last term of Eq.~(\ref{JASR}), a phase coherence term, as $\zeta=\sum_{j=1}^N e^{i{\bf k}_0\cdot {\bf r}_j}/\sqrt{N} $.

Similar to the spontaneous emission of a single atom $\hat{S}_-\hat{A}|\psi(t)\rangle=0$ and $(\langle\psi(t)|\hat{S}_+)^\dagger=\hat{S}_-|\psi(t)\rangle$ yields
 \begin{equation}
 \hat{S}_-|\psi(t)\rangle= \Big( \sum_{j=1}^N \beta_j(t) \Big)|g\rangle|0\rangle= e^{-\gamma_Nt}\:\zeta\: |g\rangle|0\rangle.
 \end{equation}
Therefore, the ensemble-WP entanglement criterion $\mu_{\scriptscriptstyle{\rm HZ}}$ becomes
\begin{equation}
\mu_{\scriptscriptstyle{\rm HZ}}^{\rm (SR)}(t) = e^{-2\gamma_Nt}\:|\zeta|^2\: |J_A(t,\gamma_N)|^2,
\end{equation}
where $J_A(t,\gamma_N)$ is given in Eq.~(\ref{JA_SE}) with $\Gamma/2\to\gamma_N$. We note that, one cannot tell if a larger $\mu_{\scriptscriptstyle{\rm HZ}}$ implies a stronger entanglement or not; neither in the WP-WP entanglement nor in ensemble-WP entanglement. This is because, unlike logarithmic negativity~\cite{plenio2005logarithmic} such entanglement criteria are not demonstrated to be used as an entanglement measure.

\section{Nonclassicality of a Wavepacket} \label{sec:WPNc}

In this section, we introduce the nonclassicality~(Nc) of a wavepacket~(WP). We show that a WP possesses nonclassicality both (a) when some of the constituent ($\bf k$) modes are squeezed or (b) when, e.g., two constituent modes ${\bf k}_1 \leftrightarrow {\bf k}_2$ are entangled. Below, we first express the two methods used for the quantification/witness of the single-mode nonclassicality~(SMNc) of a detected mode. Then, we apply these two methods for introducing the nonclassicality of a WP. 

We remind that, single-mode nonclassicality of a light mode can be defined in two different ways. (i) One may, e.g. for Gaussian states, examine the noise matrix, i.e. $V_{ij}=\langle \hat{\xi}_i \hat{\xi}_j + \hat{\xi}_j \hat{\xi}_i \rangle/2 -\langle\hat{\xi}_i\rangle \langle\hat{\xi}_j\rangle$ for the real variables $\xi^{\rm (r)}=[x_1\:,\: p_1]$ or using the complex representation $\xi^{\rm (c)}=[\alpha_1\:,\:\alpha_1^*]$~\cite{simon1994quantum,Tahira:09}. One can show that quadrature-squeezing, a SMNc, exists if $|\langle \hat{a}^2\rangle|>\langle\hat{a}^\dagger \hat{a}\rangle$~\cite{tasgin2019anatomy}, which derives from the eigenvalues of the noise matrix.

(ii) Alternatively, one can also witness/quantify the nonclassicality of a single-mode $\hat{a}$ via checking if it creates two-mode entanglement~(TME) at a BS output~\cite{Kim:02,Asboth:05,ge2015conservation}. For instance, SPH criterion~\cite{SimonPRL2000} ---not only a necessary{\&}sufficient condition for Gaussian states, but also a criterion working well for superpositions of number states--- can be used to determine the TME at the BS output. This approach may work better in witnessing the SMNc for a wider range of nonclassical states, see Fig.~2(c) in Ref.~\cite{tasgin2015measure}.

Both approaches can be used in defining the nonclassicality of a WP. We first use the method (i) to examine the states (a) and (b), expressed in the first paragraph of the present section. At the end of the section, we also mention briefly about the use of the second method (ii).

\subsection{(i) Examining the noise matrix}

Analogous to a single-mode~(SM) state, we can define the noise-matrix of a WP as
\begin{equation}
\begin{bmatrix}
\frac{1}{2}+\langle\hat{A}^\dagger\hat{A}\rangle & \langle\hat{A}^2\rangle
\\
\langle\hat{A}^2\rangle^* & \frac{1}{2}+\langle\hat{A}^\dagger\hat{A}\rangle
\end{bmatrix}
\end{equation}
in the complex representation, and as
\begin{equation}
\begin{bmatrix}
\langle\hat{X}^2\rangle-\langle\hat{X}\rangle^2 & \langle\hat{X}\hat{P}+\hat{P}\hat{X}\rangle/2-\langle\hat{X}\rangle \langle\hat{P}\rangle
\\
\langle\hat{X}\hat{P}+\hat{P}\hat{X}\rangle/2-\langle\hat{X}\rangle \langle\hat{P}\rangle & \langle\hat{P}^2\rangle-\langle\hat{P}\rangle^2
\end{bmatrix}
\end{equation}
in the real variables. Similar to SM case~\cite{tasgin2019anatomy}, $\lambda_{\rm sm}=1/2+\langle\hat{A}^\dagger\hat{A}\rangle-|\langle\hat{A}^2\rangle|$ determines the minimum noise (maximum squeezing) in the quadratures $\hat{X}_\phi=(\hat{A}_{\phi}^\dagger+\hat{A}_{\phi})/\sqrt{2}$ with  $\hat{A}_{\phi}=e^{i\phi}\hat{A}$. Here, $\phi$ is chosen along the min noise direction.

\subsubsection{(i.a) Constituent modes of a WP are squeezed}

As an example, we first examine the nonclassicality of a WP, whose some of the modes are squeezed, but the modes are all separable.

{\it Only two modes are squeezed}--- For simplicity, as a warm up, first we assume that only two modes of the WP are in squeezed vacuum state, i.e. $|\psi\rangle=|\xi_1\rangle_{{\bf k}_1}|\xi_2\rangle_{{\bf k}_2}|0\rangle_{{\bf k}_2}|0\rangle_{{\bf k}_4}\ldots$, and other modes are in vacuum state~\footnote{Actually, this is equivalent to assuming that all other modes are in coherent state. Because only the noise operators $\delta\hat{a}_i$ determine the the Nc features. $\hat{\cal D}(\alpha_i)$ displacement of each state does not alter the Nc features~\cite{simon1994quantum} for Gaussian states. \label{PS:Displacement}}. Here, $\xi_i$ are squeezed vacuum states. In such a case, only four terms non-vanish in $\langle\hat{A}^2\rangle$
\begin{eqnarray}
\langle\psi|\hat{A}^2|\psi\rangle =\langle\xi_1|\langle\xi_2|\: && \sum_{\bf r}   \sum_{{\bf r}'} \Big[ e^{i{\bf k}_1\cdot({\bf r}+\bf{r}')} \hat{a}_{{\bf k}_1}^2 + e^{i{\bf k}_2\cdot({\bf r}+\bf{r}')} \hat{a}_{{\bf k}_2}^2 
\nonumber \\
&& + 2e^{i({\bf k}_1\cdot{\bf r}+{\bf k}_2\cdot{\bf r}')} \hat{a}_{{\bf k}_1} \hat{a}_{{\bf k}_2} \Big] \: |\xi_1\rangle|\xi_2\rangle.
\label{A2k1k2_a}
\end{eqnarray}
We remark that, here, ${\bf k}_1$ and ${\bf k}_2$ are not variables, but they refer to two modes which are entangled with each other. All other modes are separable. The expectation values can be calculated by transforming the annihilation operators as $\hat{a}_i(\xi_i)=C_i\hat{a}_i-S_i\hat{a}_i^\dagger$ where $C_i\equiv \cosh r_i$ and $S_i \equiv \sinh r_i$, with $r_i$ are squeezing parameters~\cite{ScullyZubairyBook} $\xi_i=r_ie^{i\theta_i}$. We set the squeezing angles $\theta_i=0$ for simplicity.

In Eq.~(\ref{A2k1k2_a}), only $\hat{a}_{{\bf k}_i}^2$ terms survive and we obtain
\begin{equation}
\langle\psi|\hat{A}^2|\psi\rangle =-\sum_{\bf r}   \sum_{{\bf r}'} \Big( e^{i{\bf k}_1\cdot({\bf r}+\bf{r}')} S_1C_1 + e^{i{\bf k}_2\cdot({\bf r}+\bf{r}')} S_2C_2 \Big).
\label{A2k1k2}
\end{equation}
Similarly, $\langle\psi|\hat{A}^\dagger\hat{A}|\psi\rangle$ yields
\begin{equation}
\langle\psi|\hat{A}^\dagger\hat{A}|\psi\rangle =\sum_{\bf r}   \sum_{{\bf r}'} \Big( e^{i{\bf k}_1\cdot({\bf r}+\bf{r}')} S_1^2 + e^{i{\bf k}_2\cdot({\bf r}+\bf{r}')} S_2^2 \Big).
\label{ApAk1k2}
\end{equation}
One can note that  
\begin{eqnarray}
\sum_{{\bf r}'}  e^{i{\bf k}_i\cdot({\bf r}+\bf{r}')}&&= \sum_{{\bf r}'} e^{i{\bf k}_i\cdot({\bf r}-\bf{r}')}=\Big| \sum_{{\bf r}} e^{i{\bf k}_i \cdot {\bf r}} \Big|^2 \hspace{2 cm}
\nonumber \\
=&&\Big( \sum_{{\bf r}} \sin({\bf k}_i \cdot {\bf r}) \Big)^2 + \Big( \sum_{{\bf r}} \cos({\bf k}_i \cdot {\bf r})\Big)^2.
\label{sincos}
\end{eqnarray}
We remark that in the evaluation of $\langle\hat{A}^2\rangle$, in Eq.~(\ref{A2k1k2_a}), we consider only the two modes ${\bf k}_1$,${\bf k}_2$ among the summation, or $\omega$-integral, over an infinite number of modes. As could be anticipated, the contribution of the two modes remains only infinitesimal. Hence, a $|\sum_{{\bf r}} e^{i{\bf k}_i \cdot {\bf r}} |^2 $ summation, when converted to integration $|\int d^3{{\bf r}} e^{i{\bf k}_i \cdot {\bf r}}/V |^2$, vanishes. Still, we can account the infinitesimal contributions (squeezing) of the two modes to the nonclassicality of the WP as follows. $\sin({\bf k}_i \cdot {\bf r})$ summation in Eq.~(\ref{sincos}) gives exactly zero, since it is zero at ${\bf r}=0$ and symmetric/periodic terms cancel each other. In the  $\cos({\bf k}_i \cdot {\bf r})$ summation, however, the central term at ${\bf r}=0$, $\cos(0)=1$, does not vanish. Hence, following our $\sum_{\bf r}$ definition in Sec.~\ref{sec:spatial}, Eq.~(\ref{sincos}) becomes
\begin{equation}
\Big| \sum_{\bf r} e^{{\bf k}_i \cdot {\bf r}}  \Big|^2=\frac{(\Delta r)^3}{V},
\end{equation}
which is dimensionless and becomes zero in a standard continuous integration, i.e. $(\Delta r)^3/V\to 0$.

When we include this infinitesimal constribution to the noise of our WP, we obtain
\begin{eqnarray}
\lambda_{\rm sm}&&=\frac{1}{2}+\langle\hat{A}^\dagger \hat{A}\rangle - |\langle\hat{A}^2\rangle|
\nonumber \\
&&=\frac{1}{2}+\frac{(\Delta r)^3}{V} \big[(S_1^2-S_1C_1) + (S_2^2-S_2C_2)\big] ,
\end{eqnarray}
which is always less than $1/2$ since $S_i^2-S_iC_i<0$ and becomes more negative as $r_i$ increases.

{\it Many modes are squeezed}--- We are aware that, introducing the contribution from a single nonzero point, $(\Delta r)^3$ around ${\bf r}=0$, leaves an ambiguity. However, we conduct this treatment because we do need it unavoidably in the case (i,b), below. In order to leave the ambiguity, now, we also present the same treatment for a continuous distribution of the squeezing to many modes. We use the experience we obtained in our treatment with two modes.

When $|\xi_{\bf k}\rangle$ is a continuous function of $\bf k$ modes, we obtain
\begin{equation}
\langle\psi|\hat{A}^2|\psi\rangle =\langle 0|\: \sum_{{\bf r},{\bf r}'} \sum_{{\bf k},{\bf k}'} 
 e^{i{\bf k}\cdot({\bf r}+\bf{r}')} \delta_{{\bf k},{\bf k}'} \hat{a}_{\bf k}^2(\xi_{\bf k}) \:|0\rangle .
\label{A2allk}
\end{equation}
We know from Eq.~(\ref{A2k1k2_a}) that $\hat{a}_{{\bf k}} \hat{a}_{{\bf k}'}$ does not contribute. So, $\langle\psi|\hat{A}^2|\psi\rangle$ becomes
\begin{equation}
\langle\psi|\hat{A}^2|\psi\rangle = \sum_{{\bf r},{\bf r}'} \sum_{{\bf k}}
 e^{i{\bf k}\cdot({\bf r}\pm \bf{r}')} (-S_{\bf k} C_{\bf k}),
\label{A2manymode}
\end{equation}
where $S_{\bf k}\equiv \sinh r_{\bf k}$ and $C_{\bf k}\equiv \cosh r_{\bf k}$, and $r_{\bf k}$, squeezing parameter for the $\bf k$-mode, is a continuous function of $\bf k$.

If we consider a simple function, e.g. with $S_{\bf k}C_{\bf k}$ does not have any poles anywhere in the complex $\bf k$-plane, then the $\bf k$-integration in Eq.~(\ref{A2manymode}) vanishes unless ${\bf r}_1={\bf r}_2$ which leads to a single $\bf r$ summation
 \begin{equation}
\langle\psi|\hat{A}^2|\psi\rangle = \sum_{{\bf r}} \sum_{{\bf k}}
 (-S_{\bf k} C_{\bf k}) 
 =-\frac{V}{(2\pi)^3} \int d^3{\bf k} S_{\bf k} C_{\bf k},
\label{A2manymode2}
\end{equation}
where $\sum_{\bf r}=1$, see Sec.~\ref{sec:spatial}, and $\sum_{\bf k}\to \frac{V}{(2\pi)^3} \int d^3{\bf k}$ as usual~\cite{ScullyZubairyBook}. $\langle\hat{A}^\dagger \hat{A}\rangle$ can be calculated similarly as
 \begin{equation}
\langle\psi|\hat{A}^2|\psi\rangle 
 =\frac{V}{(2\pi)^3} \int d^3{\bf k} \: S_{\bf k}^2,
\label{ApAmanymode2}
\end{equation}
which gives a finite squeezing (reduction in noise)
\begin{eqnarray}
\lambda_{\rm sm}&&= \frac{1}{2} + \langle\psi|\hat{A}^\dagger\hat{A}|\psi\rangle - |\langle\psi|\hat{A}^2|\psi\rangle | 
\nonumber \\
&&= \frac{1}{2}+\frac{V}{(2\pi)^3} \int d^3{\bf k} \:(S_{\bf k}^2 - S_{\bf k}C_{\bf k})
\end{eqnarray}
for the WP. We note that $(S_{\bf k}^2 - S_{\bf k}C_{\bf k})<0$ and we remind that $S_{\bf k}\equiv \sinh r_{\bf k}$ and $C_{\bf k}\equiv \cosh r_{\bf k}$.

\subsubsection{(i.b) Entanglement of two constituent modes}

We raise the following question. Does the entanglement between two constituent modes, let them again be ${\bf k}_1$ and ${\bf k}_2$, contribute to the nonclassicality of the WP?

We consider a state, where there is no squeezing in the modes, but only the two modes ${\bf k}_1$ and ${\bf k}_2$ are entangled via two-mode squeezing operator $\hat{\tt E}=e^{\beta \hat{a}_1^\dagger\hat{a}_2^\dagger - \beta^* \hat{a}_1\hat{a}_2}$,
\begin{equation}
|\psi_{\rm ent}\rangle=|\beta\rangle_{{\bf k}_1,{\bf k}_2}\:|0\rangle_{{\bf k}_3}|0\rangle_{{\bf k}_4}\ldots 
\end{equation}
The reason we consider the entanglement due to $\hat{\tt E}$ operator is it creates "pure entanglement" between the ${\bf k}_1$ and ${\bf k}2$ modes. That is, it does create single-mode nonclassicality in the modes, see Sec. II.5.(iii) in Ref.~\cite{tasgin2019anatomy} and also Ref.~\cite{ge2015conservation}. 

We can transform the $\hat{a}_i(\beta)$ operators as
\begin{eqnarray}
\hat{a}_1(\beta)=C\hat{a}_1 + S\hat{a}_2^\dagger
\\
\hat{a}_2(\beta)=C\hat{a}_2 + S\hat{a}_1^\dagger
\end{eqnarray}
in stead of working with the entangled state $|\beta\rangle_{{\bf k}_1,{\bf k}_2}$. Here, $C\equiv\cosh r$ and $S\equiv\sinh r$ where $r$ determines the degree of the entanglement. 

In this case, only the $\hat{a}_{{\bf k}_1}(\beta) \hat{a}_{{\bf k}_2}(\beta)$ and $\hat{a}_{{\bf k}_2}(\beta) \hat{a}_{{\bf k}_1}(\beta)$ terms contribute with $CS$ in the calculation of $\langle\hat{A}^2\rangle$ and only $\hat{a}_{{\bf k}_{1,2}}^\dagger(\beta) \hat{a}_{{\bf k}_{1,2}}(\beta)$ terms contribute with $S^2$ in the calculation of $\langle\hat{A}^\dagger\hat{A}\rangle$. Thus, we find
\begin{eqnarray}
\langle\hat{A}^2\rangle_{\beta} = 2\frac{(\Delta r)^3}{V} CS,
\\
\langle\hat{A}^\dagger\hat{A}\rangle_{\beta}=2\frac{(\Delta r)^3}{V} S^2,
\end{eqnarray}
which creates an infinitesimal squeezing in the WP as
\begin{eqnarray}
\langle(\Delta \hat{X}_\phi)^2\rangle=\lambda_{\rm sm}=\frac{1}{2}+ \langle\hat{A}^\dagger\hat{A}\rangle - |\langle\hat{A}^2\rangle |
\nonumber \\
=\frac{1}{2} + 2\frac{(\Delta r)^3}{V}(S^2-SC),
\end{eqnarray}
which is always less than the SQL $1/2$. So, it creates a squeezed uncertainty WP.

\subsection{(ii) WP nonclassicality via entanglement at a beam-splitter output}

It is a known fact that single-mode nonclasicality~(SMNc) criterion $\langle\hat{a}^\dagger\hat{a}\rangle < |\langle \hat{a}^2\rangle|$, so $\langle\hat{A}^\dagger\hat{A}\rangle < |\langle \hat{A}^2\rangle|$, works good for quadrature squeezed (and Gaussian) like states. For more general states, such a nonclassicality criterion fails. In these cases, a beam-splitter~(BS) can help us very much. When a nonclassical state is input to a BS, mixed with vacuum or a coherent state, it generates two-mode entanglement~(TME) at the BS output. Hence, we can also decide that a WP is nonclassical, if it produces WP-WP entanglement at the BS output. BS transformation for a WP is given in Refs.~\cite{Zhu_BS_WP_EPL_2003}.

It is well-experienced that the SPH, two-mode entanglement, criterion~\cite{SimonPRL2000} is able to reveal the TME in some states other than the Gaussian ones, e.g. some superpositions of two-mode Fock states. Hence, determining the WP-nonclassicality via BS provides us the advantage of being able to detect some of the non-Gaussian states, e.g. superposed number states, using the strength (enhanced generality) of the SPH criterion~\footnote{SPH criterion is a strong one since it is invariant under intra-mode rotations~\cite{tasgin2019anatomy}, i.e. $\hat{a}_{1,2}=e^{i\phi_{1,2}}\hat{a}_{1,2}$.}.

For instance, use of a BS can resolve the SMNc of a superradiant-phase single-mode state, see Fig.2(c) in Ref.~\cite{tasgin2015measure}, whose nature is extremely different than the Gaussian-like states. It is a straightforward process to develop the same method, see Sec.~II.b in Ref.~\cite{tasgin2015measure}, with $\hat{a}\to\hat{A}$, also for WP-nonclassicality.

Even though SPH criterion~\cite{SimonPRL2000} is a strong one which is able to determine also some of the other states; in the Sec.~III.3 of Ref.~\cite{tasgin2019anatomy}, we developed an SPH-like (strong, invariant) criterion for number-phase squeezed like states. This new criterion is invariant under the rotations in the number-phase ($n$-$\Phi$) plane. Although SPH is a strong criterion, it is defined with quadrature variables, while the new criterion is defined with $\hat{n}$ and $\hat{\Phi}$ operators.


\section{Summary} \label{sec:summary}

Developments in the current technology necessitate entanglement/nonclassicality criteria for broadband emitting sources, e.g. like spasers~\cite{Spasers_Review2017,SpaserNanoprope2018,spaser_Entanglement_PRB_2018}. Current mode-based criteria can still be used for the broadband states. However, they detect/measure the entanglement of only between the two carrier frequencies. We introduce criteria and measures for the "total" entanglement of two wavepackets~(WPs). That is, the newly introduced criteria can measure the entanglement among all of the modes of the two WPs. Wee also develop a "total" nonclassicality for a WP, which accounts the nonclassicality of a WP both due to squeezing of the constituent modes and entanglement present among the constituent modes. In analogy with WP-WP entanglement and WP-nonclassicality, we also introduce  criteria for ensemble-WP entanglement. All the criteria/measure we introduce can also be used for measurements with near-field detectors~\cite{PlasmonDetectionNanoLett2015}.

\bibliography{bibliography}

\begin{thebibliography}{56}%
\makeatletter
\providecommand \@ifxundefined [1]{%
 \@ifx{#1\undefined}
}%
\providecommand \@ifnum [1]{%
 \ifnum #1\expandafter \@firstoftwo
 \else \expandafter \@secondoftwo
 \fi
}%
\providecommand \@ifx [1]{%
 \ifx #1\expandafter \@firstoftwo
 \else \expandafter \@secondoftwo
 \fi
}%
\providecommand \natexlab [1]{#1}%
\providecommand \enquote  [1]{``#1''}%
\providecommand \bibnamefont  [1]{#1}%
\providecommand \bibfnamefont [1]{#1}%
\providecommand \citenamefont [1]{#1}%
\providecommand \href@noop [0]{\@secondoftwo}%
\providecommand \href [0]{\begingroup \@sanitize@url \@href}%
\providecommand \@href[1]{\@@startlink{#1}\@@href}%
\providecommand \@@href[1]{\endgroup#1\@@endlink}%
\providecommand \@sanitize@url [0]{\catcode `\\12\catcode `\$12\catcode
  `\&12\catcode `\#12\catcode `\^12\catcode `\_12\catcode `\%12\relax}%
\providecommand \@@startlink[1]{}%
\providecommand \@@endlink[0]{}%
\providecommand \url  [0]{\begingroup\@sanitize@url \@url }%
\providecommand \@url [1]{\endgroup\@href {#1}{\urlprefix }}%
\providecommand \urlprefix  [0]{URL }%
\providecommand \Eprint [0]{\href }%
\providecommand \doibase [0]{http://dx.doi.org/}%
\providecommand \selectlanguage [0]{\@gobble}%
\providecommand \bibinfo  [0]{\@secondoftwo}%
\providecommand \bibfield  [0]{\@secondoftwo}%
\providecommand \translation [1]{[#1]}%
\providecommand \BibitemOpen [0]{}%
\providecommand \bibitemStop [0]{}%
\providecommand \bibitemNoStop [0]{.\EOS\space}%
\providecommand \EOS [0]{\spacefactor3000\relax}%
\providecommand \BibitemShut  [1]{\csname bibitem#1\endcsname}%
\let\auto@bib@innerbib\@empty
\bibitem [{\citenamefont {Ma}\ \emph {et~al.}(2012)\citenamefont {Ma},
  \citenamefont {Herbst}, \citenamefont {Scheidl}, \citenamefont {Wang},
  \citenamefont {Kropatschek}, \citenamefont {Naylor}, \citenamefont
  {Wittmann}, \citenamefont {Mech}, \citenamefont {Kofler}, \citenamefont
  {Anisimova} \emph {et~al.}}]{QuantumTeleportation143km}%
  \BibitemOpen
  \bibfield  {author} {\bibinfo {author} {\bibfnamefont {Xiao-Song}\
  \bibnamefont {Ma}}, \bibinfo {author} {\bibfnamefont {Thomas}\ \bibnamefont
  {Herbst}}, \bibinfo {author} {\bibfnamefont {Thomas}\ \bibnamefont
  {Scheidl}}, \bibinfo {author} {\bibfnamefont {Daqing}\ \bibnamefont {Wang}},
  \bibinfo {author} {\bibfnamefont {Sebastian}\ \bibnamefont {Kropatschek}},
  \bibinfo {author} {\bibfnamefont {William}\ \bibnamefont {Naylor}}, \bibinfo
  {author} {\bibfnamefont {Bernhard}\ \bibnamefont {Wittmann}}, \bibinfo
  {author} {\bibfnamefont {Alexandra}\ \bibnamefont {Mech}}, \bibinfo {author}
  {\bibfnamefont {Johannes}\ \bibnamefont {Kofler}}, \bibinfo {author}
  {\bibfnamefont {Elena}\ \bibnamefont {Anisimova}},  \emph {et~al.},\
  }\bibfield  {title} {\enquote {\bibinfo {title} {Quantum teleportation over
  143 kilometres using active feed-forward},}\ }\href@noop {} {\bibfield
  {journal} {\bibinfo  {journal} {Nature}\ }\textbf {\bibinfo {volume} {489}},\
  \bibinfo {pages} {269} (\bibinfo {year} {2012})}\BibitemShut {NoStop}%
\bibitem [{\citenamefont {Friis}\ \emph {et~al.}(2018)\citenamefont {Friis},
  \citenamefont {Marty}, \citenamefont {Maier}, \citenamefont {Hempel},
  \citenamefont {Holz{\"a}pfel}, \citenamefont {Jurcevic}, \citenamefont
  {Plenio}, \citenamefont {Huber}, \citenamefont {Roos}, \citenamefont {Blatt}
  \emph {et~al.}}]{20qubitsPRX2018}%
  \BibitemOpen
  \bibfield  {author} {\bibinfo {author} {\bibfnamefont {Nicolai}\ \bibnamefont
  {Friis}}, \bibinfo {author} {\bibfnamefont {Oliver}\ \bibnamefont {Marty}},
  \bibinfo {author} {\bibfnamefont {Christine}\ \bibnamefont {Maier}}, \bibinfo
  {author} {\bibfnamefont {Cornelius}\ \bibnamefont {Hempel}}, \bibinfo
  {author} {\bibfnamefont {Milan}\ \bibnamefont {Holz{\"a}pfel}}, \bibinfo
  {author} {\bibfnamefont {Petar}\ \bibnamefont {Jurcevic}}, \bibinfo {author}
  {\bibfnamefont {Martin~B}\ \bibnamefont {Plenio}}, \bibinfo {author}
  {\bibfnamefont {Marcus}\ \bibnamefont {Huber}}, \bibinfo {author}
  {\bibfnamefont {Christian}\ \bibnamefont {Roos}}, \bibinfo {author}
  {\bibfnamefont {Rainer}\ \bibnamefont {Blatt}},  \emph {et~al.},\ }\bibfield
  {title} {\enquote {\bibinfo {title} {Observation of entangled states of a
  fully controlled 20-qubit system},}\ }\href@noop {} {\bibfield  {journal}
  {\bibinfo  {journal} {Physical Review X}\ }\textbf {\bibinfo {volume} {8}},\
  \bibinfo {pages} {021012} (\bibinfo {year} {2018})}\BibitemShut {NoStop}%
\bibitem [{\citenamefont {Pirandola}\ \emph {et~al.}(2015)\citenamefont
  {Pirandola}, \citenamefont {Eisert}, \citenamefont {Weedbrook}, \citenamefont
  {Furusawa},\ and\ \citenamefont
  {Braunstein}}]{BraunsteinNatureTeleportation2015}%
  \BibitemOpen
  \bibfield  {author} {\bibinfo {author} {\bibfnamefont {S.}~\bibnamefont
  {Pirandola}}, \bibinfo {author} {\bibfnamefont {J.}~\bibnamefont {Eisert}},
  \bibinfo {author} {\bibfnamefont {C.}~\bibnamefont {Weedbrook}}, \bibinfo
  {author} {\bibfnamefont {A.}~\bibnamefont {Furusawa}}, \ and\ \bibinfo
  {author} {\bibfnamefont {S.~L.}\ \bibnamefont {Braunstein}},\ }\bibfield
  {title} {\enquote {\bibinfo {title} {Advances in quantum teleportation},}\
  }\href {https://doi.org/10.1038/nphoton.2015.154} {\bibfield  {journal}
  {\bibinfo  {journal} {Nature Photonics}\ }\textbf {\bibinfo {volume} {9}},\
  \bibinfo {pages} {641 EP --} (\bibinfo {year} {2015})},\ \bibinfo {note}
  {review Article}\BibitemShut {NoStop}%
\bibitem [{\citenamefont {Ren}\ \emph {et~al.}(2017)\citenamefont {Ren},
  \citenamefont {Xu}, \citenamefont {Yong}, \citenamefont {Zhang},
  \citenamefont {Liao}, \citenamefont {Yin}, \citenamefont {Liu}, \citenamefont
  {Cai}, \citenamefont {Yang}, \citenamefont {Li} \emph
  {et~al.}}]{SatelliteTeleportation2017}%
  \BibitemOpen
  \bibfield  {author} {\bibinfo {author} {\bibfnamefont {Ji-Gang}\ \bibnamefont
  {Ren}}, \bibinfo {author} {\bibfnamefont {Ping}\ \bibnamefont {Xu}}, \bibinfo
  {author} {\bibfnamefont {Hai-Lin}\ \bibnamefont {Yong}}, \bibinfo {author}
  {\bibfnamefont {Liang}\ \bibnamefont {Zhang}}, \bibinfo {author}
  {\bibfnamefont {Sheng-Kai}\ \bibnamefont {Liao}}, \bibinfo {author}
  {\bibfnamefont {Juan}\ \bibnamefont {Yin}}, \bibinfo {author} {\bibfnamefont
  {Wei-Yue}\ \bibnamefont {Liu}}, \bibinfo {author} {\bibfnamefont {Wen-Qi}\
  \bibnamefont {Cai}}, \bibinfo {author} {\bibfnamefont {Meng}\ \bibnamefont
  {Yang}}, \bibinfo {author} {\bibfnamefont {Li}~\bibnamefont {Li}},  \emph
  {et~al.},\ }\bibfield  {title} {\enquote {\bibinfo {title}
  {Ground-to-satellite quantum teleportation},}\ }\href@noop {} {\bibfield
  {journal} {\bibinfo  {journal} {Nature}\ }\textbf {\bibinfo {volume} {549}},\
  \bibinfo {pages} {70} (\bibinfo {year} {2017})}\BibitemShut {NoStop}%
\bibitem [{Qua()}]{QuantumRdarNews}%
  \BibitemOpen
  \href@noop {} {\enquote {\bibinfo {title} {Quantum radar:},}\ }\bibinfo
  {howpublished}
  {\url{https://www.dailymail.co.uk/news/article-6337737/China-claims-successfully-developed-QUANTUM-RADAR-detect-invisible-fighter-jets.html},
  \url{https://www.bbc.com/news/technology-43877682}},\ \bibinfo {note}
  {accessed: 2019-03-21}\BibitemShut {NoStop}%
\bibitem [{\citenamefont {Barzanjeh}\ \emph {et~al.}(2015)\citenamefont
  {Barzanjeh}, \citenamefont {Guha}, \citenamefont {Weedbrook}, \citenamefont
  {Vitali}, \citenamefont {Shapiro},\ and\ \citenamefont
  {Pirandola}}]{QuantumRadarPRL2015}%
  \BibitemOpen
  \bibfield  {author} {\bibinfo {author} {\bibfnamefont {Shabir}\ \bibnamefont
  {Barzanjeh}}, \bibinfo {author} {\bibfnamefont {Saikat}\ \bibnamefont
  {Guha}}, \bibinfo {author} {\bibfnamefont {Christian}\ \bibnamefont
  {Weedbrook}}, \bibinfo {author} {\bibfnamefont {David}\ \bibnamefont
  {Vitali}}, \bibinfo {author} {\bibfnamefont {Jeffrey~H}\ \bibnamefont
  {Shapiro}}, \ and\ \bibinfo {author} {\bibfnamefont {Stefano}\ \bibnamefont
  {Pirandola}},\ }\bibfield  {title} {\enquote {\bibinfo {title} {Microwave
  quantum illumination},}\ }\href@noop {} {\bibfield  {journal} {\bibinfo
  {journal} {Physical review letters}\ }\textbf {\bibinfo {volume} {114}},\
  \bibinfo {pages} {080503} (\bibinfo {year} {2015})}\BibitemShut {NoStop}%
\bibitem [{\citenamefont {Vaccaro}\ and\ \citenamefont
  {Pegg}(1990)}]{PeggJModOpt1990Intelligent}%
  \BibitemOpen
  \bibfield  {author} {\bibinfo {author} {\bibfnamefont {JA}~\bibnamefont
  {Vaccaro}}\ and\ \bibinfo {author} {\bibfnamefont {DT}~\bibnamefont {Pegg}},\
  }\bibfield  {title} {\enquote {\bibinfo {title} {Physical number-phase
  intelligent and minimum-uncertainty states of light},}\ }\href@noop {}
  {\bibfield  {journal} {\bibinfo  {journal} {Journal of Modern Optics}\
  }\textbf {\bibinfo {volume} {37}},\ \bibinfo {pages} {17--39} (\bibinfo
  {year} {1990})}\BibitemShut {NoStop}%
\bibitem [{\citenamefont {Duan}(2011)}]{duan2011many_particle_entanglement}%
  \BibitemOpen
  \bibfield  {author} {\bibinfo {author} {\bibfnamefont {L-M}\ \bibnamefont
  {Duan}},\ }\bibfield  {title} {\enquote {\bibinfo {title} {Entanglement
  detection in the vicinity of arbitrary dicke states},}\ }\href@noop {}
  {\bibfield  {journal} {\bibinfo  {journal} {Physical Review Letters}\
  }\textbf {\bibinfo {volume} {107}},\ \bibinfo {pages} {180502} (\bibinfo
  {year} {2011})}\BibitemShut {NoStop}%
\bibitem [{\citenamefont {Tasgin}(2017)}]{tasgin2017many}%
  \BibitemOpen
  \bibfield  {author} {\bibinfo {author} {\bibfnamefont {Mehmet~Emre}\
  \bibnamefont {Tasgin}},\ }\bibfield  {title} {\enquote {\bibinfo {title}
  {Many-particle entanglement criterion for superradiantlike states},}\
  }\href@noop {} {\bibfield  {journal} {\bibinfo  {journal} {Physical review
  letters}\ }\textbf {\bibinfo {volume} {119}},\ \bibinfo {pages} {033601}
  (\bibinfo {year} {2017})}\BibitemShut {NoStop}%
\bibitem [{\citenamefont {Bek}\ \emph {et~al.}(2006)\citenamefont {Bek},
  \citenamefont {Vogelgesang},\ and\ \citenamefont {Kern}}]{SNOMReview2006}%
  \BibitemOpen
  \bibfield  {author} {\bibinfo {author} {\bibfnamefont {Alpan}\ \bibnamefont
  {Bek}}, \bibinfo {author} {\bibfnamefont {Ralf}\ \bibnamefont {Vogelgesang}},
  \ and\ \bibinfo {author} {\bibfnamefont {Klaus}\ \bibnamefont {Kern}},\
  }\bibfield  {title} {\enquote {\bibinfo {title} {Apertureless scanning near
  field optical microscope with sub-10 nm resolution},}\ }\href@noop {}
  {\bibfield  {journal} {\bibinfo  {journal} {Review of Scientific
  Instruments}\ }\textbf {\bibinfo {volume} {77}},\ \bibinfo {pages} {043703}
  (\bibinfo {year} {2006})}\BibitemShut {NoStop}%
\bibitem [{\citenamefont {Rotenberg}\ and\ \citenamefont
  {Kuipers}(2014)}]{SNOMNature2014}%
  \BibitemOpen
  \bibfield  {author} {\bibinfo {author} {\bibfnamefont {N}~\bibnamefont
  {Rotenberg}}\ and\ \bibinfo {author} {\bibfnamefont {L}~\bibnamefont
  {Kuipers}},\ }\bibfield  {title} {\enquote {\bibinfo {title} {Mapping
  nanoscale light fields},}\ }\href@noop {} {\bibfield  {journal} {\bibinfo
  {journal} {Nature Photonics}\ }\textbf {\bibinfo {volume} {8}},\ \bibinfo
  {pages} {919} (\bibinfo {year} {2014})}\BibitemShut {NoStop}%
\bibitem [{\citenamefont {Zhang}\ \emph
  {et~al.}(2013{\natexlab{a}})\citenamefont {Zhang}, \citenamefont {Zhang},
  \citenamefont {Dong}, \citenamefont {Jiang}, \citenamefont {Zhang},
  \citenamefont {Chen}, \citenamefont {Zhang}, \citenamefont {Liao},
  \citenamefont {Aizpurua}, \citenamefont {Luo} \emph
  {et~al.}}]{SERS_subnmNature2013}%
  \BibitemOpen
  \bibfield  {author} {\bibinfo {author} {\bibfnamefont {Renhe}\ \bibnamefont
  {Zhang}}, \bibinfo {author} {\bibfnamefont {Y}~\bibnamefont {Zhang}},
  \bibinfo {author} {\bibfnamefont {ZC}~\bibnamefont {Dong}}, \bibinfo {author}
  {\bibfnamefont {S}~\bibnamefont {Jiang}}, \bibinfo {author} {\bibfnamefont
  {C}~\bibnamefont {Zhang}}, \bibinfo {author} {\bibfnamefont {LG}~\bibnamefont
  {Chen}}, \bibinfo {author} {\bibfnamefont {L}~\bibnamefont {Zhang}}, \bibinfo
  {author} {\bibfnamefont {Y}~\bibnamefont {Liao}}, \bibinfo {author}
  {\bibfnamefont {J}~\bibnamefont {Aizpurua}}, \bibinfo {author} {\bibfnamefont
  {Y~ea}\ \bibnamefont {Luo}},  \emph {et~al.},\ }\bibfield  {title} {\enquote
  {\bibinfo {title} {Chemical mapping of a single molecule by plasmon-enhanced
  raman scattering},}\ }\href@noop {} {\bibfield  {journal} {\bibinfo
  {journal} {Nature}\ }\textbf {\bibinfo {volume} {498}},\ \bibinfo {pages}
  {82} (\bibinfo {year} {2013}{\natexlab{a}})}\BibitemShut {NoStop}%
\bibitem [{\citenamefont {Zong}\ \emph {et~al.}(2018)\citenamefont {Zong},
  \citenamefont {Xu}, \citenamefont {Xu}, \citenamefont {Wei}, \citenamefont
  {Ma}, \citenamefont {Zheng}, \citenamefont {Hu},\ and\ \citenamefont
  {Ren}}]{SERS_bioanalysis_Review_NanoLett_2018}%
  \BibitemOpen
  \bibfield  {author} {\bibinfo {author} {\bibfnamefont {Cheng}\ \bibnamefont
  {Zong}}, \bibinfo {author} {\bibfnamefont {Mengxi}\ \bibnamefont {Xu}},
  \bibinfo {author} {\bibfnamefont {Li-Jia}\ \bibnamefont {Xu}}, \bibinfo
  {author} {\bibfnamefont {Ting}\ \bibnamefont {Wei}}, \bibinfo {author}
  {\bibfnamefont {Xin}\ \bibnamefont {Ma}}, \bibinfo {author} {\bibfnamefont
  {Xiao-Shan}\ \bibnamefont {Zheng}}, \bibinfo {author} {\bibfnamefont {Ren}\
  \bibnamefont {Hu}}, \ and\ \bibinfo {author} {\bibfnamefont {Bin}\
  \bibnamefont {Ren}},\ }\bibfield  {title} {\enquote {\bibinfo {title}
  {Surface-enhanced raman spectroscopy for bioanalysis: reliability and
  challenges},}\ }\href@noop {} {\bibfield  {journal} {\bibinfo  {journal}
  {Chemical reviews}\ }\textbf {\bibinfo {volume} {118}},\ \bibinfo {pages}
  {4946--4980} (\bibinfo {year} {2018})}\BibitemShut {NoStop}%
\bibitem [{\citenamefont {Scully}\ and\ \citenamefont
  {Zubairy}(1997)}]{ScullyZubairyBook}%
  \BibitemOpen
  \bibfield  {author} {\bibinfo {author} {\bibfnamefont {M.~O.}\ \bibnamefont
  {Scully}}\ and\ \bibinfo {author} {\bibfnamefont {M.~S.}\ \bibnamefont
  {Zubairy}},\ }\href@noop {} {\emph {\bibinfo {title} {Quantum Optics}}}\
  (\bibinfo  {publisher} {Cambridge University Press},\ \bibinfo {address} {New
  York},\ \bibinfo {year} {1997})\BibitemShut {NoStop}%
\bibitem [{\citenamefont {Limonov}\ \emph {et~al.}(2017)\citenamefont
  {Limonov}, \citenamefont {Rybin}, \citenamefont {Poddubny},\ and\
  \citenamefont {Kivshar}}]{Fano_resonances_in_photonics_NaturePhot_2017}%
  \BibitemOpen
  \bibfield  {author} {\bibinfo {author} {\bibfnamefont {Mikhail~F}\
  \bibnamefont {Limonov}}, \bibinfo {author} {\bibfnamefont {Mikhail~V}\
  \bibnamefont {Rybin}}, \bibinfo {author} {\bibfnamefont {Alexander~N}\
  \bibnamefont {Poddubny}}, \ and\ \bibinfo {author} {\bibfnamefont {Yuri~S}\
  \bibnamefont {Kivshar}},\ }\bibfield  {title} {\enquote {\bibinfo {title}
  {Fano resonances in photonics},}\ }\href@noop {} {\bibfield  {journal}
  {\bibinfo  {journal} {Nature Photonics}\ }\textbf {\bibinfo {volume} {11}},\
  \bibinfo {pages} {543} (\bibinfo {year} {2017})}\BibitemShut {NoStop}%
\bibitem [{\citenamefont
  {Stockman}(2010{\natexlab{a}})}]{stockman_Nature_2010_Dark_hot_resonances}%
  \BibitemOpen
  \bibfield  {author} {\bibinfo {author} {\bibfnamefont {Mark~I}\ \bibnamefont
  {Stockman}},\ }\bibfield  {title} {\enquote {\bibinfo {title} {Nanoscience:
  Dark-hot resonances},}\ }\href@noop {} {\bibfield  {journal} {\bibinfo
  {journal} {Nature}\ }\textbf {\bibinfo {volume} {467}},\ \bibinfo {pages}
  {541} (\bibinfo {year} {2010}{\natexlab{a}})}\BibitemShut {NoStop}%
\bibitem [{\citenamefont {Ta{\c{s}}g{\i}n}\ \emph {et~al.}(2018)\citenamefont
  {Ta{\c{s}}g{\i}n}, \citenamefont {Bek},\ and\ \citenamefont
  {Postac{\i}}}]{TasginFanoBook2018}%
  \BibitemOpen
  \bibfield  {author} {\bibinfo {author} {\bibfnamefont {Mehmet~Emre}\
  \bibnamefont {Ta{\c{s}}g{\i}n}}, \bibinfo {author} {\bibfnamefont {Alpan}\
  \bibnamefont {Bek}}, \ and\ \bibinfo {author} {\bibfnamefont {Selen}\
  \bibnamefont {Postac{\i}}},\ }\bibfield  {title} {\enquote {\bibinfo {title}
  {Fano resonances in the linear and nonlinear plasmonic response},}\ }in\
  \href@noop {} {\emph {\bibinfo {booktitle} {Fano Resonances in Optics and
  Microwaves}}}\ (\bibinfo  {publisher} {Springer},\ \bibinfo {year} {2018})\
  pp.\ \bibinfo {pages} {1--31}\BibitemShut {NoStop}%
\bibitem [{\citenamefont {Zhang}\ \emph
  {et~al.}(2013{\natexlab{b}})\citenamefont {Zhang}, \citenamefont {Wen},
  \citenamefont {Zhen}, \citenamefont {Nordlander},\ and\ \citenamefont
  {Halas}}]{PNAS_2013_Fano_FWM}%
  \BibitemOpen
  \bibfield  {author} {\bibinfo {author} {\bibfnamefont {Yu}~\bibnamefont
  {Zhang}}, \bibinfo {author} {\bibfnamefont {Fangfang}\ \bibnamefont {Wen}},
  \bibinfo {author} {\bibfnamefont {Yu-Rong}\ \bibnamefont {Zhen}}, \bibinfo
  {author} {\bibfnamefont {Peter}\ \bibnamefont {Nordlander}}, \ and\ \bibinfo
  {author} {\bibfnamefont {Naomi~J}\ \bibnamefont {Halas}},\ }\bibfield
  {title} {\enquote {\bibinfo {title} {Coherent fano resonances in a plasmonic
  nanocluster enhance optical four-wave mixing},}\ }\href@noop {} {\bibfield
  {journal} {\bibinfo  {journal} {Proceedings of the National Academy of
  Sciences}\ }\textbf {\bibinfo {volume} {110}},\ \bibinfo {pages} {9215--9219}
  (\bibinfo {year} {2013}{\natexlab{b}})}\BibitemShut {NoStop}%
\bibitem [{\citenamefont {He}\ \emph {et~al.}(2016)\citenamefont {He},
  \citenamefont {Fan}, \citenamefont {Ding}, \citenamefont {Zhu},\ and\
  \citenamefont {Liang}}]{SciRep_2016_Fano_CARS}%
  \BibitemOpen
  \bibfield  {author} {\bibinfo {author} {\bibfnamefont {Jinna}\ \bibnamefont
  {He}}, \bibinfo {author} {\bibfnamefont {Chunzhen}\ \bibnamefont {Fan}},
  \bibinfo {author} {\bibfnamefont {Pei}\ \bibnamefont {Ding}}, \bibinfo
  {author} {\bibfnamefont {Shuangmei}\ \bibnamefont {Zhu}}, \ and\ \bibinfo
  {author} {\bibfnamefont {Erjun}\ \bibnamefont {Liang}},\ }\bibfield  {title}
  {\enquote {\bibinfo {title} {Near-field engineering of fano resonances in a
  plasmonic assembly for maximizing cars enhancements},}\ }\href@noop {}
  {\bibfield  {journal} {\bibinfo  {journal} {Scientific reports}\ }\textbf
  {\bibinfo {volume} {6}},\ \bibinfo {pages} {20777} (\bibinfo {year}
  {2016})}\BibitemShut {NoStop}%
\bibitem [{\citenamefont {Postaci}\ \emph {et~al.}(2018)\citenamefont
  {Postaci}, \citenamefont {Yildiz}, \citenamefont {Bek},\ and\ \citenamefont
  {Tasgin}}]{tasgin_Nanophotonics_2018_SilentSERS}%
  \BibitemOpen
  \bibfield  {author} {\bibinfo {author} {\bibfnamefont {Selen}\ \bibnamefont
  {Postaci}}, \bibinfo {author} {\bibfnamefont {Bilge~Can}\ \bibnamefont
  {Yildiz}}, \bibinfo {author} {\bibfnamefont {Alpan}\ \bibnamefont {Bek}}, \
  and\ \bibinfo {author} {\bibfnamefont {Mehmet~Emre}\ \bibnamefont {Tasgin}},\
  }\bibfield  {title} {\enquote {\bibinfo {title} {Silent enhancement of sers
  signal without increasing hot spot intensities},}\ }\href@noop {} {\bibfield
  {journal} {\bibinfo  {journal} {Nanophotonics}\ }\textbf {\bibinfo {volume}
  {7}},\ \bibinfo {pages} {1687--1695} (\bibinfo {year} {2018})}\BibitemShut
  {NoStop}%
\bibitem [{\citenamefont {Wu}\ \emph {et~al.}(2010)\citenamefont {Wu},
  \citenamefont {Gray},\ and\ \citenamefont {Pelton}}]{PeltonOptExp2010}%
  \BibitemOpen
  \bibfield  {author} {\bibinfo {author} {\bibfnamefont {Xiaohua}\ \bibnamefont
  {Wu}}, \bibinfo {author} {\bibfnamefont {Stephen~K}\ \bibnamefont {Gray}}, \
  and\ \bibinfo {author} {\bibfnamefont {Matthew}\ \bibnamefont {Pelton}},\
  }\bibfield  {title} {\enquote {\bibinfo {title} {Quantum-dot-induced
  transparency in a nanoscale plasmonic resonator},}\ }\href@noop {} {\bibfield
   {journal} {\bibinfo  {journal} {Optics express}\ }\textbf {\bibinfo {volume}
  {18}},\ \bibinfo {pages} {23633--23645} (\bibinfo {year} {2010})}\BibitemShut
  {NoStop}%
\bibitem [{\citenamefont {Kosionis}\ \emph {et~al.}(2012)\citenamefont
  {Kosionis}, \citenamefont {Terzis}, \citenamefont {Sadeghi},\ and\
  \citenamefont {Paspalakis}}]{PaspalakisJPhys2012}%
  \BibitemOpen
  \bibfield  {author} {\bibinfo {author} {\bibfnamefont {Spyridon~G}\
  \bibnamefont {Kosionis}}, \bibinfo {author} {\bibfnamefont {Andreas~F}\
  \bibnamefont {Terzis}}, \bibinfo {author} {\bibfnamefont {Seyed~M}\
  \bibnamefont {Sadeghi}}, \ and\ \bibinfo {author} {\bibfnamefont {Emmanuel}\
  \bibnamefont {Paspalakis}},\ }\bibfield  {title} {\enquote {\bibinfo {title}
  {Optical response of a quantum dot--metal nanoparticle hybrid interacting
  with a weak probe field},}\ }\href@noop {} {\bibfield  {journal} {\bibinfo
  {journal} {Journal of Physics: Condensed Matter}\ }\textbf {\bibinfo {volume}
  {25}},\ \bibinfo {pages} {045304} (\bibinfo {year} {2012})}\BibitemShut
  {NoStop}%
\bibitem [{\citenamefont {Di~Martino}\ \emph {et~al.}(2012)\citenamefont
  {Di~Martino}, \citenamefont {Sonnefraud}, \citenamefont {K{\'e}na-Cohen},
  \citenamefont {Tame}, \citenamefont {Ozdemir}, \citenamefont {Kim},\ and\
  \citenamefont {Maier}}]{PlasmonEntanglementLifetime_NanoLett_2012}%
  \BibitemOpen
  \bibfield  {author} {\bibinfo {author} {\bibfnamefont {Giuliana}\
  \bibnamefont {Di~Martino}}, \bibinfo {author} {\bibfnamefont {Yannick}\
  \bibnamefont {Sonnefraud}}, \bibinfo {author} {\bibfnamefont {St{\'e}phane}\
  \bibnamefont {K{\'e}na-Cohen}}, \bibinfo {author} {\bibfnamefont {Mark}\
  \bibnamefont {Tame}}, \bibinfo {author} {\bibfnamefont {Sahin~K}\
  \bibnamefont {Ozdemir}}, \bibinfo {author} {\bibfnamefont {MS}~\bibnamefont
  {Kim}}, \ and\ \bibinfo {author} {\bibfnamefont {Stefan~A}\ \bibnamefont
  {Maier}},\ }\bibfield  {title} {\enquote {\bibinfo {title} {Quantum
  statistics of surface plasmon polaritons in metallic stripe waveguides},}\
  }\href@noop {} {\bibfield  {journal} {\bibinfo  {journal} {Nano letters}\
  }\textbf {\bibinfo {volume} {12}},\ \bibinfo {pages} {2504--2508} (\bibinfo
  {year} {2012})}\BibitemShut {NoStop}%
\bibitem [{\citenamefont {Huck}\ \emph {et~al.}(2009)\citenamefont {Huck},
  \citenamefont {Smolka}, \citenamefont {Lodahl}, \citenamefont {S{\o}rensen},
  \citenamefont {Boltasseva}, \citenamefont {Janousek},\ and\ \citenamefont
  {Andersen}}]{SqueezedPlasmons_PRL_2009}%
  \BibitemOpen
  \bibfield  {author} {\bibinfo {author} {\bibfnamefont {Alexander}\
  \bibnamefont {Huck}}, \bibinfo {author} {\bibfnamefont {Stephan}\
  \bibnamefont {Smolka}}, \bibinfo {author} {\bibfnamefont {Peter}\
  \bibnamefont {Lodahl}}, \bibinfo {author} {\bibfnamefont {Anders~S}\
  \bibnamefont {S{\o}rensen}}, \bibinfo {author} {\bibfnamefont {Alexandra}\
  \bibnamefont {Boltasseva}}, \bibinfo {author} {\bibfnamefont {Jiri}\
  \bibnamefont {Janousek}}, \ and\ \bibinfo {author} {\bibfnamefont {Ulrik~L}\
  \bibnamefont {Andersen}},\ }\bibfield  {title} {\enquote {\bibinfo {title}
  {Demonstration of quadrature-squeezed surface plasmons in a gold
  waveguide},}\ }\href@noop {} {\bibfield  {journal} {\bibinfo  {journal}
  {Physical review letters}\ }\textbf {\bibinfo {volume} {102}},\ \bibinfo
  {pages} {246802} (\bibinfo {year} {2009})}\BibitemShut {NoStop}%
\bibitem [{\citenamefont {Altewischer}\ \emph {et~al.}(2002)\citenamefont
  {Altewischer}, \citenamefont {Van~Exter},\ and\ \citenamefont
  {Woerdman}}]{altewischer2002plasmon}%
  \BibitemOpen
  \bibfield  {author} {\bibinfo {author} {\bibfnamefont {E}~\bibnamefont
  {Altewischer}}, \bibinfo {author} {\bibfnamefont {MP}~\bibnamefont
  {Van~Exter}}, \ and\ \bibinfo {author} {\bibfnamefont {JP}~\bibnamefont
  {Woerdman}},\ }\bibfield  {title} {\enquote {\bibinfo {title}
  {Plasmon-assisted transmission of entangled photons},}\ }\href@noop {}
  {\bibfield  {journal} {\bibinfo  {journal} {Nature}\ }\textbf {\bibinfo
  {volume} {418}},\ \bibinfo {pages} {304} (\bibinfo {year}
  {2002})}\BibitemShut {NoStop}%
\bibitem [{\citenamefont {Chen}\ and\ \citenamefont
  {Chen}(2012)}]{EntanglementFanoOptLett2012}%
  \BibitemOpen
  \bibfield  {author} {\bibinfo {author} {\bibfnamefont {Guang-Yin}\
  \bibnamefont {Chen}}\ and\ \bibinfo {author} {\bibfnamefont {Yueh-Nan}\
  \bibnamefont {Chen}},\ }\bibfield  {title} {\enquote {\bibinfo {title}
  {Correspondence between entanglement and fano resonance of surface
  plasmons},}\ }\href@noop {} {\bibfield  {journal} {\bibinfo  {journal}
  {Optics letters}\ }\textbf {\bibinfo {volume} {37}},\ \bibinfo {pages}
  {4023--4025} (\bibinfo {year} {2012})}\BibitemShut {NoStop}%
\bibitem [{\citenamefont {Dong}\ \emph {et~al.}(2015)\citenamefont {Dong},
  \citenamefont {Zhang}, \citenamefont {Zheng},\ and\ \citenamefont
  {Sun}}]{FluorescenceEnhancementNanophotonics2015}%
  \BibitemOpen
  \bibfield  {author} {\bibinfo {author} {\bibfnamefont {Jun}\ \bibnamefont
  {Dong}}, \bibinfo {author} {\bibfnamefont {Zhenglong}\ \bibnamefont {Zhang}},
  \bibinfo {author} {\bibfnamefont {Hairong}\ \bibnamefont {Zheng}}, \ and\
  \bibinfo {author} {\bibfnamefont {Mentao}\ \bibnamefont {Sun}},\ }\bibfield
  {title} {\enquote {\bibinfo {title} {Recent progress on plasmon-enhanced
  fluorescence},}\ }\href@noop {} {\bibfield  {journal} {\bibinfo  {journal}
  {Nanophotonics}\ }\textbf {\bibinfo {volume} {4}},\ \bibinfo {pages}
  {472--490} (\bibinfo {year} {2015})}\BibitemShut {NoStop}%
\bibitem [{\citenamefont {Wang}\ \emph {et~al.}(2017)\citenamefont {Wang},
  \citenamefont {Meng}, \citenamefont {Kildishev}, \citenamefont {Boltasseva},\
  and\ \citenamefont {Shalaev}}]{Spasers_Review2017}%
  \BibitemOpen
  \bibfield  {author} {\bibinfo {author} {\bibfnamefont {Zhuoxian}\
  \bibnamefont {Wang}}, \bibinfo {author} {\bibfnamefont {Xiangeng}\
  \bibnamefont {Meng}}, \bibinfo {author} {\bibfnamefont {Alexander~V}\
  \bibnamefont {Kildishev}}, \bibinfo {author} {\bibfnamefont {Alexandra}\
  \bibnamefont {Boltasseva}}, \ and\ \bibinfo {author} {\bibfnamefont
  {Vladimir~M}\ \bibnamefont {Shalaev}},\ }\bibfield  {title} {\enquote
  {\bibinfo {title} {Nanolasers enabled by metallic nanoparticles: from spasers
  to random lasers},}\ }\href@noop {} {\bibfield  {journal} {\bibinfo
  {journal} {Laser \& Photonics Reviews}\ }\textbf {\bibinfo {volume} {11}},\
  \bibinfo {pages} {1700212} (\bibinfo {year} {2017})}\BibitemShut {NoStop}%
\bibitem [{\citenamefont {Song}\ \emph {et~al.}(2018)\citenamefont {Song},
  \citenamefont {Wang}, \citenamefont {Zhang}, \citenamefont {Yang},
  \citenamefont {Lu}, \citenamefont {Kang}, \citenamefont {Xu},\ and\
  \citenamefont {Chen}}]{SpaserNanoprope2018}%
  \BibitemOpen
  \bibfield  {author} {\bibinfo {author} {\bibfnamefont {Pei}\ \bibnamefont
  {Song}}, \bibinfo {author} {\bibfnamefont {Jian-Hua}\ \bibnamefont {Wang}},
  \bibinfo {author} {\bibfnamefont {Miao}\ \bibnamefont {Zhang}}, \bibinfo
  {author} {\bibfnamefont {Fan}\ \bibnamefont {Yang}}, \bibinfo {author}
  {\bibfnamefont {Hai-Jie}\ \bibnamefont {Lu}}, \bibinfo {author}
  {\bibfnamefont {Bin}\ \bibnamefont {Kang}}, \bibinfo {author} {\bibfnamefont
  {Jing-Juan}\ \bibnamefont {Xu}}, \ and\ \bibinfo {author} {\bibfnamefont
  {Hong-Yuan}\ \bibnamefont {Chen}},\ }\bibfield  {title} {\enquote {\bibinfo
  {title} {Three-level spaser for next-generation luminescent nanoprobe},}\
  }\href@noop {} {\bibfield  {journal} {\bibinfo  {journal} {Science advances}\
  }\textbf {\bibinfo {volume} {4}},\ \bibinfo {pages} {eaat0292} (\bibinfo
  {year} {2018})}\BibitemShut {NoStop}%
\bibitem [{\citenamefont {Noginov}\ \emph {et~al.}(2009)\citenamefont
  {Noginov}, \citenamefont {Zhu}, \citenamefont {Belgrave}, \citenamefont
  {Bakker}, \citenamefont {Shalaev}, \citenamefont {Narimanov}, \citenamefont
  {Stout}, \citenamefont {Herz}, \citenamefont {Suteewong},\ and\ \citenamefont
  {Wiesner}}]{noginov2009demonstration}%
  \BibitemOpen
  \bibfield  {author} {\bibinfo {author} {\bibfnamefont {MA}~\bibnamefont
  {Noginov}}, \bibinfo {author} {\bibfnamefont {G}~\bibnamefont {Zhu}},
  \bibinfo {author} {\bibfnamefont {AM}~\bibnamefont {Belgrave}}, \bibinfo
  {author} {\bibfnamefont {Reuben}\ \bibnamefont {Bakker}}, \bibinfo {author}
  {\bibfnamefont {VM}~\bibnamefont {Shalaev}}, \bibinfo {author} {\bibfnamefont
  {EE}~\bibnamefont {Narimanov}}, \bibinfo {author} {\bibfnamefont
  {S}~\bibnamefont {Stout}}, \bibinfo {author} {\bibfnamefont {E}~\bibnamefont
  {Herz}}, \bibinfo {author} {\bibfnamefont {T}~\bibnamefont {Suteewong}}, \
  and\ \bibinfo {author} {\bibfnamefont {U}~\bibnamefont {Wiesner}},\
  }\bibfield  {title} {\enquote {\bibinfo {title} {Demonstration of a
  spaser-based nanolaser},}\ }\href@noop {} {\bibfield  {journal} {\bibinfo
  {journal} {Nature}\ }\textbf {\bibinfo {volume} {460}},\ \bibinfo {pages}
  {1110} (\bibinfo {year} {2009})}\BibitemShut {NoStop}%
\bibitem [{\citenamefont {Stockman}(2010{\natexlab{b}})}]{stockman2010spaser}%
  \BibitemOpen
  \bibfield  {author} {\bibinfo {author} {\bibfnamefont {Mark~I}\ \bibnamefont
  {Stockman}},\ }\bibfield  {title} {\enquote {\bibinfo {title} {The spaser as
  a nanoscale quantum generator and ultrafast amplifier},}\ }\href@noop {}
  {\bibfield  {journal} {\bibinfo  {journal} {Journal of Optics}\ }\textbf
  {\bibinfo {volume} {12}},\ \bibinfo {pages} {024004} (\bibinfo {year}
  {2010}{\natexlab{b}})}\BibitemShut {NoStop}%
\bibitem [{\citenamefont {Da{\u{g}}}\ \emph {et~al.}(2016)\citenamefont
  {Da{\u{g}}}, \citenamefont {Niedenzu}, \citenamefont
  {M{\"u}stecapl{\i}o{\u{g}}lu},\ and\ \citenamefont
  {Kurizki}}]{MustecapQHeatEngine2016}%
  \BibitemOpen
  \bibfield  {author} {\bibinfo {author} {\bibfnamefont {Ceren}\ \bibnamefont
  {Da{\u{g}}}}, \bibinfo {author} {\bibfnamefont {Wolfgang}\ \bibnamefont
  {Niedenzu}}, \bibinfo {author} {\bibfnamefont {{\"O}zg{\"u}r}\ \bibnamefont
  {M{\"u}stecapl{\i}o{\u{g}}lu}}, \ and\ \bibinfo {author} {\bibfnamefont
  {Gershon}\ \bibnamefont {Kurizki}},\ }\bibfield  {title} {\enquote {\bibinfo
  {title} {Multiatom quantum coherences in micromasers as fuel for thermal and
  nonthermal machines},}\ }\href@noop {} {\bibfield  {journal} {\bibinfo
  {journal} {Entropy}\ }\textbf {\bibinfo {volume} {18}},\ \bibinfo {pages}
  {244} (\bibinfo {year} {2016})}\BibitemShut {NoStop}%
\bibitem [{\citenamefont {Da{\u{g}}}\ \emph {et~al.}(2018)\citenamefont
  {Da{\u{g}}}, \citenamefont {Niedenzu}, \citenamefont {Ozaydin}, \citenamefont
  {M{\"u}stecapl{\i}o{\u{g}}lu},\ and\ \citenamefont
  {Kurizki}}]{MustecapQHeatEngine2018}%
  \BibitemOpen
  \bibfield  {author} {\bibinfo {author} {\bibfnamefont {Ceren~B}\ \bibnamefont
  {Da{\u{g}}}}, \bibinfo {author} {\bibfnamefont {Wolfgang}\ \bibnamefont
  {Niedenzu}}, \bibinfo {author} {\bibfnamefont {Fatih}\ \bibnamefont
  {Ozaydin}}, \bibinfo {author} {\bibfnamefont {{\"O}zg{\"u}r~E}\ \bibnamefont
  {M{\"u}stecapl{\i}o{\u{g}}lu}}, \ and\ \bibinfo {author} {\bibfnamefont
  {Gershon}\ \bibnamefont {Kurizki}},\ }\bibfield  {title} {\enquote {\bibinfo
  {title} {Temperature control in dissipative cavities by entangled dimers},}\
  }\href@noop {} {\bibfield  {journal} {\bibinfo  {journal} {The Journal of
  Physical Chemistry C}\ } (\bibinfo {year} {2018})}\BibitemShut {NoStop}%
\bibitem [{\citenamefont {Goodfellow}\ \emph {et~al.}(2015)\citenamefont
  {Goodfellow}, \citenamefont {Chakraborty}, \citenamefont {Beams},
  \citenamefont {Novotny},\ and\ \citenamefont
  {Vamivakas}}]{PlasmonDetectionNanoLett2015}%
  \BibitemOpen
  \bibfield  {author} {\bibinfo {author} {\bibfnamefont {Kenneth~M}\
  \bibnamefont {Goodfellow}}, \bibinfo {author} {\bibfnamefont {Chitraleema}\
  \bibnamefont {Chakraborty}}, \bibinfo {author} {\bibfnamefont {Ryan}\
  \bibnamefont {Beams}}, \bibinfo {author} {\bibfnamefont {Lukas}\ \bibnamefont
  {Novotny}}, \ and\ \bibinfo {author} {\bibfnamefont {A~Nick}\ \bibnamefont
  {Vamivakas}},\ }\bibfield  {title} {\enquote {\bibinfo {title} {Direct
  on-chip optical plasmon detection with an atomically thin semiconductor},}\
  }\href@noop {} {\bibfield  {journal} {\bibinfo  {journal} {Nano letters}\
  }\textbf {\bibinfo {volume} {15}},\ \bibinfo {pages} {5477--5481} (\bibinfo
  {year} {2015})}\BibitemShut {NoStop}%
\bibitem [{\citenamefont {Gubin}\ \emph {et~al.}(2018)\citenamefont {Gubin},
  \citenamefont {Shesterikov}, \citenamefont {Karpov},\ and\ \citenamefont
  {Prokhorov}}]{spaser_Entanglement_PRB_2018}%
  \BibitemOpen
  \bibfield  {author} {\bibinfo {author} {\bibfnamefont {M~Yu}\ \bibnamefont
  {Gubin}}, \bibinfo {author} {\bibfnamefont {AV}~\bibnamefont {Shesterikov}},
  \bibinfo {author} {\bibfnamefont {SN}~\bibnamefont {Karpov}}, \ and\ \bibinfo
  {author} {\bibfnamefont {AV}~\bibnamefont {Prokhorov}},\ }\bibfield  {title}
  {\enquote {\bibinfo {title} {Entangled plasmon generation in nonlinear spaser
  system under the action of external magnetic field},}\ }\href@noop {}
  {\bibfield  {journal} {\bibinfo  {journal} {Physical Review B}\ }\textbf
  {\bibinfo {volume} {97}},\ \bibinfo {pages} {085431} (\bibinfo {year}
  {2018})}\BibitemShut {NoStop}%
\bibitem [{\citenamefont {Julsgaard}\ \emph {et~al.}(2001)\citenamefont
  {Julsgaard}, \citenamefont {Kozhekin},\ and\ \citenamefont
  {Polzik}}]{PolzikNature2001ensemble}%
  \BibitemOpen
  \bibfield  {author} {\bibinfo {author} {\bibfnamefont {Brian}\ \bibnamefont
  {Julsgaard}}, \bibinfo {author} {\bibfnamefont {Alexander}\ \bibnamefont
  {Kozhekin}}, \ and\ \bibinfo {author} {\bibfnamefont {Eugene~S}\ \bibnamefont
  {Polzik}},\ }\bibfield  {title} {\enquote {\bibinfo {title} {Experimental
  long-lived entanglement of two macroscopic objects},}\ }\href@noop {}
  {\bibfield  {journal} {\bibinfo  {journal} {Nature}\ }\textbf {\bibinfo
  {volume} {413}},\ \bibinfo {pages} {400--403} (\bibinfo {year}
  {2001})}\BibitemShut {NoStop}%
\bibitem [{\citenamefont {Qurban}\ \emph {et~al.}(2018)\citenamefont {Qurban},
  \citenamefont {Ikram}, \citenamefont {Ge},\ and\ \citenamefont
  {Zubairy}}]{ZubairyQDs_Plasmon_entanglement}%
  \BibitemOpen
  \bibfield  {author} {\bibinfo {author} {\bibfnamefont {Misbah}\ \bibnamefont
  {Qurban}}, \bibinfo {author} {\bibfnamefont {Manzoor}\ \bibnamefont {Ikram}},
  \bibinfo {author} {\bibfnamefont {Guo-Qin}\ \bibnamefont {Ge}}, \ and\
  \bibinfo {author} {\bibfnamefont {M~Suhail}\ \bibnamefont {Zubairy}},\
  }\bibfield  {title} {\enquote {\bibinfo {title} {Entanglement generation
  among quantum dots and surface plasmons of a metallic nanoring},}\
  }\href@noop {} {\bibfield  {journal} {\bibinfo  {journal} {Journal of Physics
  B: Atomic, Molecular and Optical Physics}\ }\textbf {\bibinfo {volume}
  {51}},\ \bibinfo {pages} {155502} (\bibinfo {year} {2018})}\BibitemShut
  {NoStop}%
\bibitem [{\citenamefont {Simon}(2000)}]{SimonPRL2000}%
  \BibitemOpen
  \bibfield  {author} {\bibinfo {author} {\bibfnamefont {R.}~\bibnamefont
  {Simon}},\ }\bibfield  {title} {\enquote {\bibinfo {title} {Peres-horodecki
  separability criterion for continuous variable systems},}\ }\href {\doibase
  10.1103/PhysRevLett.84.2726} {\bibfield  {journal} {\bibinfo  {journal}
  {Phys. Rev. Lett.}\ }\textbf {\bibinfo {volume} {84}},\ \bibinfo {pages}
  {2726--2729} (\bibinfo {year} {2000})}\BibitemShut {NoStop}%
\bibitem [{\citenamefont {Duan}\ \emph {et~al.}(2000)\citenamefont {Duan},
  \citenamefont {Giedke}, \citenamefont {Cirac},\ and\ \citenamefont
  {Zoller}}]{DGCZ_PRL2000}%
  \BibitemOpen
  \bibfield  {author} {\bibinfo {author} {\bibfnamefont {Lu-Ming}\ \bibnamefont
  {Duan}}, \bibinfo {author} {\bibfnamefont {G.}~\bibnamefont {Giedke}},
  \bibinfo {author} {\bibfnamefont {J.}~\bibnamefont {Cirac}}, \ and\ \bibinfo
  {author} {\bibfnamefont {P.}~\bibnamefont {Zoller}},\ }\bibfield  {title}
  {\enquote {\bibinfo {title} {Inseparability criterion for continuous variable
  systems},}\ }\href {\doibase 10.1103/PhysRevLett.84.2722} {\bibfield
  {journal} {\bibinfo  {journal} {Phys. Rev. Lett.}\ }\textbf {\bibinfo
  {volume} {84}},\ \bibinfo {pages} {2722--2725} (\bibinfo {year}
  {2000})}\BibitemShut {NoStop}%
\bibitem [{\citenamefont {Hillery}\ and\ \citenamefont
  {Zubairy}(2006)}]{Hillery&ZubairyPRL2006}%
  \BibitemOpen
  \bibfield  {author} {\bibinfo {author} {\bibfnamefont {Mark}\ \bibnamefont
  {Hillery}}\ and\ \bibinfo {author} {\bibfnamefont {M.}~\bibnamefont
  {Zubairy}},\ }\bibfield  {title} {\enquote {\bibinfo {title} {Entanglement
  conditions for two-mode states},}\ }\href {\doibase
  10.1103/PhysRevLett.96.050503} {\bibfield  {journal} {\bibinfo  {journal}
  {Phys. Rev. Lett.}\ }\textbf {\bibinfo {volume} {96}},\ \bibinfo {pages}
  {050503} (\bibinfo {year} {2006})}\BibitemShut {NoStop}%
\bibitem [{\citenamefont {Shchukin}\ and\ \citenamefont
  {Vogel}(2005)}]{ShchukinVogelPRL2005}%
  \BibitemOpen
  \bibfield  {author} {\bibinfo {author} {\bibfnamefont {E}~\bibnamefont
  {Shchukin}}\ and\ \bibinfo {author} {\bibfnamefont {W}~\bibnamefont
  {Vogel}},\ }\bibfield  {title} {\enquote {\bibinfo {title} {Inseparability
  criteria for continuous bipartite quantum states},}\ }\href@noop {}
  {\bibfield  {journal} {\bibinfo  {journal} {Physical Review Letters}\
  }\textbf {\bibinfo {volume} {95}},\ \bibinfo {pages} {230502} (\bibinfo
  {year} {2005})}\BibitemShut {NoStop}%
\bibitem [{\citenamefont {Nha}\ and\ \citenamefont
  {Kim}(2006)}]{NhaPRA2006Fock_states}%
  \BibitemOpen
  \bibfield  {author} {\bibinfo {author} {\bibfnamefont {Hyunchul}\
  \bibnamefont {Nha}}\ and\ \bibinfo {author} {\bibfnamefont {Jaewan}\
  \bibnamefont {Kim}},\ }\bibfield  {title} {\enquote {\bibinfo {title}
  {Entanglement criteria via the uncertainty relations in su (2) and su (1, 1)
  algebras: Detection of non-gaussian entangled states},}\ }\href@noop {}
  {\bibfield  {journal} {\bibinfo  {journal} {Physical Review A}\ }\textbf
  {\bibinfo {volume} {74}},\ \bibinfo {pages} {012317} (\bibinfo {year}
  {2006})}\BibitemShut {NoStop}%
\bibitem [{\citenamefont {L\'opez}\ \emph {et~al.}(2008)\citenamefont
  {L\'opez}, \citenamefont {Romero}, \citenamefont {Lastra}, \citenamefont
  {Solano},\ and\ \citenamefont {Retamal}}]{SunndenBirthofEntang_PRL_2008}%
  \BibitemOpen
  \bibfield  {author} {\bibinfo {author} {\bibfnamefont {C.~E.}\ \bibnamefont
  {L\'opez}}, \bibinfo {author} {\bibfnamefont {G.}~\bibnamefont {Romero}},
  \bibinfo {author} {\bibfnamefont {F.}~\bibnamefont {Lastra}}, \bibinfo
  {author} {\bibfnamefont {E.}~\bibnamefont {Solano}}, \ and\ \bibinfo {author}
  {\bibfnamefont {J.~C.}\ \bibnamefont {Retamal}},\ }\bibfield  {title}
  {\enquote {\bibinfo {title} {Sudden birth versus sudden death of entanglement
  in multipartite systems},}\ }\href {\doibase 10.1103/PhysRevLett.101.080503}
  {\bibfield  {journal} {\bibinfo  {journal} {Phys. Rev. Lett.}\ }\textbf
  {\bibinfo {volume} {101}},\ \bibinfo {pages} {080503} (\bibinfo {year}
  {2008})}\BibitemShut {NoStop}%
\bibitem [{\citenamefont {Moore}\ and\ \citenamefont
  {Meystre}(1999)}]{MeystrePRL1999_BEC_SR}%
  \BibitemOpen
  \bibfield  {author} {\bibinfo {author} {\bibfnamefont {MG}~\bibnamefont
  {Moore}}\ and\ \bibinfo {author} {\bibfnamefont {Pierre}\ \bibnamefont
  {Meystre}},\ }\bibfield  {title} {\enquote {\bibinfo {title} {Theory of
  superradiant scattering of laser light from bose-einstein condensates},}\
  }\href@noop {} {\bibfield  {journal} {\bibinfo  {journal} {Physical Review
  Letters}\ }\textbf {\bibinfo {volume} {83}},\ \bibinfo {pages} {5202}
  (\bibinfo {year} {1999})}\BibitemShut {NoStop}%
\bibitem [{\citenamefont {Ta{\c{s}}g{\i}n}\ \emph {et~al.}(2011)\citenamefont
  {Ta{\c{s}}g{\i}n}, \citenamefont {M{\"u}stecapl{\i}og̃lu},\ and\
  \citenamefont {You}}]{tasginPRA2011vortex}%
  \BibitemOpen
  \bibfield  {author} {\bibinfo {author} {\bibfnamefont {Mehmet~Emre}\
  \bibnamefont {Ta{\c{s}}g{\i}n}}, \bibinfo {author} {\bibfnamefont
  {{\"O}E}~\bibnamefont {M{\"u}stecapl{\i}og̃lu}}, \ and\ \bibinfo {author}
  {\bibfnamefont {L}~\bibnamefont {You}},\ }\bibfield  {title} {\enquote
  {\bibinfo {title} {Creation of a vortex in a bose-einstein condensate by
  superradiant scattering},}\ }\href@noop {} {\bibfield  {journal} {\bibinfo
  {journal} {Physical Review A}\ }\textbf {\bibinfo {volume} {84}},\ \bibinfo
  {pages} {063628} (\bibinfo {year} {2011})}\BibitemShut {NoStop}%
\bibitem [{\citenamefont {Tasgin}(2019)}]{tasgin2019anatomy}%
  \BibitemOpen
  \bibfield  {author} {\bibinfo {author} {\bibfnamefont {Mehmet~Emre}\
  \bibnamefont {Tasgin}},\ }\bibfield  {title} {\enquote {\bibinfo {title}
  {Anatomy of entanglement and nonclassicality criteria},}\ }\href@noop {}
  {\bibfield  {journal} {\bibinfo  {journal} {arXiv preprint arXiv:1901.04045}\
  } (\bibinfo {year} {2019})}\BibitemShut {NoStop}%
\bibitem [{\citenamefont {Svidzinsky}\ and\ \citenamefont
  {Chang}(2008)}]{svidzinskyPRA2008cooperative}%
  \BibitemOpen
  \bibfield  {author} {\bibinfo {author} {\bibfnamefont {Anatoly}\ \bibnamefont
  {Svidzinsky}}\ and\ \bibinfo {author} {\bibfnamefont {Jun-Tao}\ \bibnamefont
  {Chang}},\ }\bibfield  {title} {\enquote {\bibinfo {title} {Cooperative
  spontaneous emission as a many-body eigenvalue problem},}\ }\href@noop {}
  {\bibfield  {journal} {\bibinfo  {journal} {Physical Review A}\ }\textbf
  {\bibinfo {volume} {77}},\ \bibinfo {pages} {043833} (\bibinfo {year}
  {2008})}\BibitemShut {NoStop}%
\bibitem [{\citenamefont {Scully}\ and\ \citenamefont
  {Svidzinsky}(2009)}]{ScullyScience2009SR}%
  \BibitemOpen
  \bibfield  {author} {\bibinfo {author} {\bibfnamefont {Marlan~O}\
  \bibnamefont {Scully}}\ and\ \bibinfo {author} {\bibfnamefont {Anatoly~A}\
  \bibnamefont {Svidzinsky}},\ }\bibfield  {title} {\enquote {\bibinfo {title}
  {The super of superradiance},}\ }\href@noop {} {\bibfield  {journal}
  {\bibinfo  {journal} {Science}\ }\textbf {\bibinfo {volume} {325}},\ \bibinfo
  {pages} {1510--1511} (\bibinfo {year} {2009})}\BibitemShut {NoStop}%
\bibitem [{\citenamefont {Plenio}(2005)}]{plenio2005logarithmic}%
  \BibitemOpen
  \bibfield  {author} {\bibinfo {author} {\bibfnamefont {Martin~B}\
  \bibnamefont {Plenio}},\ }\bibfield  {title} {\enquote {\bibinfo {title}
  {Logarithmic negativity: A full entanglement monotone that is not convex},}\
  }\href@noop {} {\bibfield  {journal} {\bibinfo  {journal} {Phys. Rev. Lett.}\
  }\textbf {\bibinfo {volume} {95}},\ \bibinfo {pages} {090503} (\bibinfo
  {year} {2005})}\BibitemShut {NoStop}%
\bibitem [{\citenamefont {Simon}\ \emph {et~al.}(1994)\citenamefont {Simon},
  \citenamefont {Mukunda},\ and\ \citenamefont {Dutta}}]{simon1994quantum}%
  \BibitemOpen
  \bibfield  {author} {\bibinfo {author} {\bibfnamefont {R}~\bibnamefont
  {Simon}}, \bibinfo {author} {\bibfnamefont {N}~\bibnamefont {Mukunda}}, \
  and\ \bibinfo {author} {\bibfnamefont {Biswadeb}\ \bibnamefont {Dutta}},\
  }\bibfield  {title} {\enquote {\bibinfo {title} {Quantum-noise matrix for
  multimode systems: U (n) invariance, squeezing, and normal forms},}\
  }\href@noop {} {\bibfield  {journal} {\bibinfo  {journal} {Physical Review
  A}\ }\textbf {\bibinfo {volume} {49}},\ \bibinfo {pages} {1567} (\bibinfo
  {year} {1994})}\BibitemShut {NoStop}%
\bibitem [{\citenamefont {Tahira}\ \emph {et~al.}(2009)\citenamefont {Tahira},
  \citenamefont {Ikram}, \citenamefont {Nha},\ and\ \citenamefont
  {Zubairy}}]{Tahira:09}%
  \BibitemOpen
  \bibfield  {author} {\bibinfo {author} {\bibfnamefont {Rabia}\ \bibnamefont
  {Tahira}}, \bibinfo {author} {\bibfnamefont {Manzoor}\ \bibnamefont {Ikram}},
  \bibinfo {author} {\bibfnamefont {Hyunchul}\ \bibnamefont {Nha}}, \ and\
  \bibinfo {author} {\bibfnamefont {M.~Suhail}\ \bibnamefont {Zubairy}},\
  }\bibfield  {title} {\enquote {\bibinfo {title} {Entanglement of gaussian
  states using a beam splitter},}\ }\href {\doibase 10.1103/PhysRevA.79.023816}
  {\bibfield  {journal} {\bibinfo  {journal} {Phys. Rev. A}\ }\textbf {\bibinfo
  {volume} {79}},\ \bibinfo {pages} {023816} (\bibinfo {year}
  {2009})}\BibitemShut {NoStop}%
\bibitem [{\citenamefont {Kim}\ \emph {et~al.}(2002)\citenamefont {Kim},
  \citenamefont {Son}, \citenamefont {Bu\ifmmode~\check{z}\else \v{z}\fi{}ek},\
  and\ \citenamefont {Knight}}]{Kim:02}%
  \BibitemOpen
  \bibfield  {author} {\bibinfo {author} {\bibfnamefont {M.~S.}\ \bibnamefont
  {Kim}}, \bibinfo {author} {\bibfnamefont {W.}~\bibnamefont {Son}}, \bibinfo
  {author} {\bibfnamefont {V.}~\bibnamefont {Bu\ifmmode~\check{z}\else
  \v{z}\fi{}ek}}, \ and\ \bibinfo {author} {\bibfnamefont {P.~L.}\ \bibnamefont
  {Knight}},\ }\bibfield  {title} {\enquote {\bibinfo {title} {Entanglement by
  a beam splitter: Nonclassicality as a prerequisite for entanglement},}\
  }\href {\doibase 10.1103/PhysRevA.65.032323} {\bibfield  {journal} {\bibinfo
  {journal} {Phys. Rev. A}\ }\textbf {\bibinfo {volume} {65}},\ \bibinfo
  {pages} {032323} (\bibinfo {year} {2002})}\BibitemShut {NoStop}%
\bibitem [{\citenamefont {Asb\'oth}\ \emph {et~al.}(2005)\citenamefont
  {Asb\'oth}, \citenamefont {Calsamiglia},\ and\ \citenamefont
  {Ritsch}}]{Asboth:05}%
  \BibitemOpen
  \bibfield  {author} {\bibinfo {author} {\bibfnamefont {J\'anos~K.}\
  \bibnamefont {Asb\'oth}}, \bibinfo {author} {\bibfnamefont {John}\
  \bibnamefont {Calsamiglia}}, \ and\ \bibinfo {author} {\bibfnamefont
  {Helmut}\ \bibnamefont {Ritsch}},\ }\bibfield  {title} {\enquote {\bibinfo
  {title} {Computable measure of nonclassicality for light},}\ }\href {\doibase
  10.1103/PhysRevLett.94.173602} {\bibfield  {journal} {\bibinfo  {journal}
  {Phys. Rev. Lett.}\ }\textbf {\bibinfo {volume} {94}},\ \bibinfo {pages}
  {173602} (\bibinfo {year} {2005})}\BibitemShut {NoStop}%
\bibitem [{\citenamefont {Ge}\ \emph {et~al.}(2015)\citenamefont {Ge},
  \citenamefont {Tasgin},\ and\ \citenamefont {Zubairy}}]{ge2015conservation}%
  \BibitemOpen
  \bibfield  {author} {\bibinfo {author} {\bibfnamefont {Wenchao}\ \bibnamefont
  {Ge}}, \bibinfo {author} {\bibfnamefont {Mehmet~Emre}\ \bibnamefont
  {Tasgin}}, \ and\ \bibinfo {author} {\bibfnamefont {M~Suhail}\ \bibnamefont
  {Zubairy}},\ }\bibfield  {title} {\enquote {\bibinfo {title} {Conservation
  relation of nonclassicality and entanglement for gaussian states in a beam
  splitter},}\ }\href@noop {} {\bibfield  {journal} {\bibinfo  {journal}
  {Physical Review A}\ }\textbf {\bibinfo {volume} {92}},\ \bibinfo {pages}
  {052328} (\bibinfo {year} {2015})}\BibitemShut {NoStop}%
\bibitem [{\citenamefont {Tasgin}(2015)}]{tasgin2015measure}%
  \BibitemOpen
  \bibfield  {author} {\bibinfo {author} {\bibfnamefont {Mehmet~Emre}\
  \bibnamefont {Tasgin}},\ }\bibfield  {title} {\enquote {\bibinfo {title}
  {Single-mode nonclassicality measure from simon-peres-horodecki criterion},}\
  }\href@noop {} {\bibfield  {journal} {\bibinfo  {journal} {arXiv preprint
  arXiv:1502.00992}\ } (\bibinfo {year} {2015})}\BibitemShut {NoStop}%
\bibitem [{\citenamefont {Wang}\ and\ \citenamefont
  {Zhu}(2003)}]{Zhu_BS_WP_EPL_2003}%
  \BibitemOpen
  \bibfield  {author} {\bibinfo {author} {\bibfnamefont {Kaige}\ \bibnamefont
  {Wang}}\ and\ \bibinfo {author} {\bibfnamefont {Shiyao}\ \bibnamefont
  {Zhu}},\ }\bibfield  {title} {\enquote {\bibinfo {title} {Two-photon
  anti-coalescence interference: The signature of two-photon entanglement},}\
  }\href {\doibase 10.1209/epl/i2003-00503-6} {\bibfield  {journal} {\bibinfo
  {journal} {Europhysics Letters ({EPL})}\ }\textbf {\bibinfo {volume} {64}},\
  \bibinfo {pages} {22--28} (\bibinfo {year} {2003})}\BibitemShut {NoStop}%
\end{thebibliography}%
  
\end{document}